\documentclass[11pt,a4paper]{article}
\usepackage{jheppub}

\usepackage{verbatim, mathrsfs, array, layout, textcomp, latexsym, slashed, booktabs, mathtools, tikz, dsfont, comment, enumerate, multirow, amsthm, xspace, physics, commath, cleveref,subfigure}

\numberwithin{equation}{section}

\title{Broken Conformal Window}

\author[1]{Dan Kondo}
\author[1,2,3]{Hitoshi Murayama}
\author[2,3]{Bea Noether}
\author[2,3]{Digvijay Roy Varier}

\affiliation[1]{Kavli Institute for the Physics and Mathematics of the
  Universe (WPI), University of Tokyo, Kashiwa 277-8583, Japan}

\affiliation[2]{Department of Physics, University of California, Berkeley, CA 94720, USA}

\affiliation[3]{Theoretical Physics Group, Lawrence Berkeley National Laboratory,
Berkeley, CA 94720, USA}

\emailAdd{dan.kondo@ipmu.jp}
\emailAdd{hitoshi@berkeley.edu}
\emailAdd{hitoshi.murayama@ipmu.jp}
\emailAdd{bea\_noether@berkeley.edu}
\emailAdd{digvijayroyvarier@berkeley.edu}

\abstract{
We show that near the edges of the conformal window of supersymmetric $SU(N_c)$ QCD, perturbed by Anomaly Mediated Supersymmetry Breaking (AMSB), chiral symmetry can be broken depending on the initial conditions of the RG flow. We do so by perturbatively expanding around Banks--Zaks fixed points and taking advantage of Seiberg duality. Interpolating between the edges of the conformal window, we predict that non-supersymmetric QCD breaks chiral symmetry up to $N_f \leq 3N_c - 1$, while we cannot say anything definitive for $N_f \geq 3N_c$ at this moment.
}

\begin{document}
\maketitle
\section{Introduction}
 While it is true that this world is made of simple blocks like elementary particles or atoms, a plethora of possibilities arise from many body or strongly correlated systems due to the presence of interactions between diverse sets of these particles. Such systems exhibit fascinating and complicated behaviors beyond simple non-interacting models, but are difficult to study in general. They appear in high-$T_c$ superconductors, heavy fermion systems, and frustrated spin systems in condensed matter physics~\cite{avella2014strongly}, and quark confinement, chiral symmetry breaking~\cite{greensite2011introduction,rothe2012lattice}, and nuclear binding in nuclear physics~\cite{bohr1998nuclear,ring2004nuclear}. In all of these cases, the fundamental Hamiltonian or Lagrangian is believed to be ``known,'' while it is theoretically difficult to work out their consequences at long distances or infrared (IR) limits. Such systems cannot be studied within perturbation theory because we do not have a good candidate for the ``zeroth order approximation'' that captures the qualitative feature of the system. Once we have such a ``zeroth order approximation'', we can identify how to approach a realistic system using perturbation. 

Nuclear physics is believed to be a consequence of Quantum Chromo-Dynamics (QCD), a non-abelian gauge theory based on the $SU(3)$ gauge group, with quarks and gluons
as fundamental fields. It is known that QCD is an asymptotically-free theory, {\it i.e.}\/, the coupling becomes weak at short distances but strong at long distances \cite{Gross:1973id,Politzer:1973fx}. It is also known that in nature, quarks have masses. Therefore, chiral symmetry cannot be an exact symmetry of the QCD Lagrangian. Yet, it is difficult to understand how exactly this dynamical chiral symmetry breaking arises from the fundamental Lagrangian for QCD with massless quarks \cite{Nambu:1961tp,Nambu:1961fr}. It is responsible for making the pion mass light and allowing nuclei to bind beyond the size of individual nucleons. Path integrals over fields can be performed on supercomputers (\emph{ e.g.}, lattice QCD), but they are limited in that the quark masses cannot be taken to zero. 

It was proposed~\cite{Murayama:2021xfj} that the supersymmetric version of QCD (SQCD), which had been studied in seminal work by Seiberg \cite{Seiberg:1994bz,Seiberg:1994pq}, combined with anomaly-mediated supersymmetry breaking (AMSB) \cite{Randall:1998uk,Giudice:1998xp}, provides an effective tool for studying the dynamics of QCD beyond perturbation theory. This tool may well be applicable beyond QCD, and in fact it led to new predictions on the IR dynamics of chiral gauge theories \cite{Csaki:2021xhi,Csaki:2021aqv,Kondo:2022lvu} different from past conjectures. Furthermore, the application to $SO(N_c)$ gauge theories led to the demonstration that both magnetic monopoles and mesons condensed in the ground state, practically a proof of confinement in these theories \cite{Csaki:2021jax,Csaki:2021xuc}. It is also expected that the tool can be applied to lower dimensional systems such as for strongly correlated systems in condensed matter physics. 

However, questions remained about the so-called ``conformal window,'', where the SQCD dynamics leads to an infrared (IR) fixed point for $SU(N_c)$ gauge theories with a number of flavors within the range $\frac{3}{2}N_c < N_f < 3N_c$. Such theories have been shown to be superconformal in the long-distance limit \cite{Seiberg:1994pq}. Once perturbed by AMSB, it was immediately clear that the AMSB effects disappear asymptotically in the fixed point limit. However, it was not clear whether the theories reach IR fixed points in the presence of such a perturbation. 

In this paper, we argue that the AMSB effects are relevant and deflect the renormalization-group-equation (RGE) flow to chiral symmetry breaking under certain conditions. We demonstrate that the question can be studied at the upper and lower edges of the conformal window for large $N_c$, where IR fixed points are achieved within the validity of perturbation theory. The AMSB effects are shown to be relevant in these regimes, and therefore, the dynamics does not lead to superconformal fixed points. Instead, we show a convincing picture that the theories lead to chiral symmetry breaking if the superpartners of quarks get positive mass and can be decoupled. 

We predict that non-supersymmetric QCD leads to chiral symmetry breaking all the way up to $N_f < 3 N_c$ 
Our analysis cannot be applied to the range $N_f \geq 3 N_c$ because the AMSB effects make the squark masses negative without a ground state in SQCD for $N_f \geq 3N_c$. 

The paper is organized as follows. We first briefly review the features of anomaly mediated supersymmetry breaking (AMSB) in ~\Cref{sec:AMSB}, and the notion of conformal windows in non-Abelian gauge theories in ~\Cref{sec:ConformalWindow}. In~\Cref{sec:RGEoriginalelectric}, we calculate the RGE in the electric theory and show that AMSB can be relevant, for the magnetic dual description in~\Cref{sec:RGEmagnetic}, and for the ``twice-dual" description in~\Cref{sec:RGEelectricTwiceDual}. In~\Cref{sec:VacuumelectricTwicedual}, we derive a chiral symmetry broken ($\chi$SB) vacuum configuration in the twice-dual description under certain conditions, and for the magnetic theory in~\Cref{sec:Vacuummagnetic}. In~\Cref{sec:allASQCD}, we summarize the AMSB predictions regarding $\chi$SB vacua in non-SUSY QCD for all ranges of $N_f$ and $N_c$ before concluding in~\Cref{sec:Conclusion}.

\section{Anomaly Mediation}\label{sec:AMSB}

Here, we very briefly review Anomaly mediation of supersymmetry breaking (AMSB)~\cite{Randall:1998uk,Giudice:1998xp}, which can be formulated with the Weyl compensator ${ \Phi} = 1 + \theta^{2} m$~\cite{Pomarol:1999ie} that appears in the supersymmetric Lagrangian as
\begin{align}
	{\cal L}&=\int d^{4}\theta\Phi^{*}\Phi K 
    +\int d^{2}\theta\Phi^{3}W + c.c.
\end{align}
Here, $K$ is the K\"ahler potential and $W$ is the superpotential of the theory, and $m$ is the supersymmetry breaking parameter. When the theory is conformal, ${\Phi} $ can be removed from the theory by rescaling the fields $\phi_{i} \rightarrow {\Phi} ^{-1} \phi_{i}$. On the other hand, violation of conformal invariance leads to supersymmetry breaking effects. Solving for auxiliary fields, the superpotential leads to the tree-level supersymmetry breaking terms
\begin{align}
	{V}_{\rm tree} &= m \left( \phi_{i} \frac{\partial W}{\partial \phi_{i}} - 3 W \right)
	+ c.c.
	\label{eq:tree}
\end{align}
Dimensionless coupling constants do not lead to supersymmetry breaking effects because of the conformal invariance at tree-level. However, conformal invariance is anomalously broken due to the running of coupling constants, and there are loop-level supersymmetry breaking effects in tri-linear couplings, scalar masses, and gaugino masses,
\begin{align}
	A_{ijk} (\mu) &=  -\frac{1}{2} (\gamma_{i} + \gamma_{j} + \gamma_{k})(\mu) m, 
	\label{eq:A} \\
	m_{i}^{2}(\mu) &=  -\frac{1}{4} \dot{\gamma}_{i}(\mu) m^{2}, 
	\label{eq:m2} \\
	m_{\lambda}(\mu) &= - \frac{\beta_g}{2g^{2}}(\mu) m. \label{eq:mlambda}
\end{align}
where $\gamma_i = \mu\dod{}{\mu}\ln Z_i(\mu)$ and $\dot{\gamma}_i = \mu\dod{}{\mu}\gamma_i$.\\
In the case of a superpotential $W = \frac{1}{3!} \lambda_{ijk} \phi_i \phi_j \phi_k$, at one loop one has
\begin{align}
\gamma_{i} =  \frac{1}{8\pi^2} (2g^2 C_i - \frac{1}{2} \sum_{j,k}\lambda_{ijk}^*\lambda_{ijk})
\end{align}
Here, $C_i = T^a T^a$ is the quadratic Casimir for the representation $i$. Note that we follow the notation $\beta_g= \mu \frac{d}{d\mu} g^{2}$ etc. In general, physical masses are the sum of contributions from the superpotential (tree-level or non-perturbative), tree-level AMSB~\cref{eq:tree} and loop-level AMSB~\cref{eq:A,eq:m2,eq:mlambda}.

The most remarkable property of AMSB is its ultraviolet-insensitivity. The expressions for the supersymmetry breaking parameters above depend on wave function renormalization and running coupling constants, which jump when heavy fields are integrated out from the theory. It turns out that the threshold corrections from the loops of heavy fields precisely give the necessary jump. Therefore the above expressions remain true at {\it all}\/ energy scales and depend only on the particle content and interactions present at that energy scale. This point can be verified explicitly in perturbative calculations, and is very transparent in the $\overline{\rm DR}$ scheme~\cite{Boyda:2001nh}. This means, for example, that we can determine the IR dynamics of an AMSB-perturbed theory by perturbing the effective IR description of the SUSY theory.\\

For later use, let us derive the potential for given K\"ahler potential $K$ and superpotential $W$, which reproduce the result in~\cite{Csaki:2022cyg}
\begin{align}\label{eq:AMSBpotential}
V_{ \text{tree} } &= \partial_i Wg^{ij*} \partial^*_j W^*+m^*m \left( \partial_i Kg^{ij*}\partial^*_jK-K \right)
+m (\partial_iWg^{ij*}\partial^*_jK-3W)+ \text{c.c},
\end{align}
where the index $i$ ($j*$) specifies the (anti-)chiral superfield, $g_{ij*}=\partial_i\partial^*_jK$ is the K\"ahler metric and $g^{ij*}$ is the inverse of $g_{ij*}$.\\
The SUSY breaking from anomaly mediation comes from the superconformal anomaly in a supergravity (SUGRA) background. Its effects can be efficiently obtained by rescaling operators with the Weyl compensator field $\Phi=1+m\theta^2$ according to their conformal weight~\cite{Pomarol:1999ie}. The tree level SUGRA Lagrangian with the Weyl compensator is
\begin{align}\label{eq:TreelevelLagrangianAMSB}
\mathcal{L} &= -3 \left[ \mathrm{e}^{-\frac{K}{3}}\Phi\Phi^\dag \right]_D + \left[ W\Phi^3 \right]_F + (\text{c.c})\nonumber\\
&= -3 \left[ \mathrm{e}^{-\frac{K}{3}}  (1+m\theta^2)(1+m\bar{\theta}^2) \right]_D + \left[ W  (1+3m\theta^2) \right]_F \nonumber\\
&\simeq  \left[ \left(K+\frac{\partial K}{\partial\phi_i}F_i\theta^2+ \frac{\partial K}{\partial \bar{\phi}_i}\bar{F}_i\bar{\theta}^2+\frac{\partial^2K}{\partial \phi_i \partial\bar{\phi}_j} F_i\bar{F}_j \theta^2\bar{\theta}^2\right)  (1+m\theta^2)(1+m\bar{\theta}^2) \right]_D \nonumber\\
&+ \left[ \left(W +\frac{\partial W}{\partial \phi_i} F_i\theta^2\right)  (1+3m\theta^2) \right]_F+ \text{c.c} \nonumber\\
&=
m^2K+m(K_iF_i+K_{\bar{i}} \bar{F}_i) +K_{i\bar{j}} F_i\bar{F}_j +3mW+3mW^*+W_i F_i+W^*_{\bar{i}}\bar{F}_i \nonumber
\end{align}
The equation of motion of $F$ and $\bar{F}$ are
\begin{align}
mK_i+K_{i\bar{j}} \bar{F}_j +W_i &=0 \\
mK_{\bar{i}} +K_{l\bar{i}} F_l + W^*_{\bar{i}} &=0 
\end{align}
this can be solved with inverse K\"ahler metric
\begin{align}
F_i &=-m (K^{-1} )^{\bar{l}}_ i K_{\bar{l}} -  (K^{-1} )^{\bar{l}}_i W^*_{\bar{l}} \\
\bar{F}_{\bar{j}} &= -m(K^{-1})^i_{\bar{j}} K_i - (K^{-1})^i_{\bar{j}} W_i
\end{align}
Substituting these back into Lagrangian~\cref{eq:TreelevelLagrangianAMSB} reads
\begin{align}\label{eq:AMSBLagrangiantouse}
\mathcal{L} 
&\simeq
m^2K+m(K_iF_i+K_{\bar{i}} \bar{F}_i) +K_{i\bar{j}} F_i\bar{F}_j +3mW+3mW^*+W_i F_i+W^*_{\bar{i}}\bar{F}_i  \nonumber\\
& \nonumber \\
&=
m^2K-m(K_i(m (K^{-1} )^{\bar{l}}_ i K_{\bar{l}} + (K^{-1} )^{\bar{l}}_i W^*_{\bar{l}} )+K_{\bar{i}} ( m(K^{-1})^i_{\bar{j}} K_i +(K^{-1})^i_{\bar{j}} W_i )) \nonumber\\
& +K_{i\bar{j}} ( -m (K^{-1} )^{\bar{l}}_ i K_{\bar{l}} -  (K^{-1} )^{\bar{l}}_i W^*_{\bar{l}} )  (  -m(K^{-1})^i_{\bar{j}} K_i - (K^{-1})^i_{\bar{j}} W_i ) +3mW \nonumber\\
&+3mW^*-W_i (m (K^{-1} )^{\bar{l}}_ i K_{\bar{l}} +  (K^{-1} )^{\bar{l}}_i W^*_{\bar{l}})  +W^*_{\bar{i}}  ( m(K^{-1})^i_{\bar{j}} K_i +(K^{-1})^i_{\bar{j}} W_i )   \nonumber\\ 
& \nonumber \\
&=
m^2(K -K_{\bar{i}} g^{m\bar{i}} K_m) 
+m(3W+3W^* 
-W_i  g^{i\bar{l}} K_{\bar{l}}  -W^*_{\bar{i}}g^{i\bar{j}} K_i) -W^*_{\bar{i}}g^{i\bar{j}} W_i.
\end{align}
They enter the potential through $V\supset -\mathcal{L}$, and this is the potential~\cref{eq:AMSBpotential}. It is important to remember that the K\"ahler potential can include the wave function renormalization factor
\begin{align}\label{eq:wavefunctionR}
    \mathcal{Z}_i=Z_i\left[1-\frac{\gamma_i}{2}m^2(\theta^2+\bar{\theta}^2)+\frac{\gamma_i^2+\dot{\gamma}_i}{4}m^2\theta^2\bar{\theta}^2\right],
\end{align}
which serves as a superfield spurion that encodes the loop-level AMSB effects. Typically these effects are sub-dominant, but near a superconformal phase the anomalous dimensions become large and these effects can be relevant, as we will see below.

\section{The Conformal Window}\label{sec:ConformalWindow}
Generally speaking, in non-Abelian gauge theory with multiple flavors $N_f$ of massless fermions with representation $R$, there is a conformal window (for review, see e.g.~\cite{Nogradi:2016qek}). In the region $N_f^{\rm{I}\hspace{-1pt}\rm{I}}<N_f<N_f^{\rm{I}}$, the theory is asymptotically free in the short distance physics, while the theory is scale invariant, governed by a non-trivial fixed point in the long distance physics. The upper limit of that range, $N_f^{\rm{I}}$ is determined by perturbation theory. According to the $\beta$-function
\begin{align}
    \beta(x)&=\frac{dx}{d\log\mu^2}=-(\beta_0x^2+\beta_1x^3+\cdots),
\end{align}
where $x=\alpha_s/\pi$. In $N_c=3$ QCD, for example, the upper bound comes from the change in the sign of the first term indicating the loss of asymptotic freedom for $N_f^{\rm{I}}>16.5$. For $N_f<N_f^{\rm{I}}$, $\beta_0>0$ while $\beta_1<0$, and this opposite sign leads to the non-trivial fixed point $x_{\text{FP}}=-\beta_0/\beta_1$. The smallness of the coupling makes it easier to determine the upper bound $N_f^{\rm{I}}$. On the contrary, it is difficult to determine the lower bound $N_f^{\rm{I}\hspace{-1pt}\rm{I}}$, below which confinement or chiral symmetry breaking is thought to set in.

Estimated values of $N_f^{\rm{I}\hspace{-1pt}\rm{I}}$ for $SU(3)$ from various approaches are within $8<N_f^{\rm{I}\hspace{-1pt}\rm{I}}<13$~\cite{Appelquist:1988yc,Cohen:1988sq,Sannino:2004qp,Dietrich:2006cm,Armoni:2009jn,Braun:2009ns,Frandsen:2010ej,Rychkov_2017,Kim:2020yvr,Lee:2020ihn}. Whether $N_f=12$ is in the conformal window or not has attracted recent lattice studies~\cite{Appelquist:2007hu,Appelquist:2009ty,Fodor:2009wk,Fodor:2011tu,Hasenfratz:2011xn,DeGrand:2011cu,Lin:2012iw,Aoki:2012eq,LatKMI:2013bhp,Fodor:2016zil,Hasenfratz:2016dou,Mickley:2025mjj}. Some studies indicate that the chiral symmetry broken ($\chi$SB) vacuum is realized up to and possibly beyond $N_f=3N_c$. For instance, one paper finds that $SU(2)$ gauge theories exhibit chiral symmetry breaking for $N_f < 6$~\cite{Amato:2018nvj}, while another finds that $SU(3)$ gauge theory with $N_f=8$ breaks chiral symmetry~\cite{LatKMI:2014xoh}, and another finds that $SU(3)$ with $N_f=12$ appears to reach an IR fixed point~\cite{LatKMI:2013bhp}. These results are still subject to discussion ~\cite{DeGrand:2015zxa}.

Now, let us think about this question from the approach of SUSY-QCD (SQCD). Seiberg established the conformal window of SQCD for $\frac{3}{2} N_c < N_f < 3 N_c$, where the theory flows to IR fixed points with non-trivial superconformal dynamics~\cite{Seiberg:1994pq,Seiberg:1994bz,Intriligator:1995au}. The $SU(N_c)$ electric theory ($N_f$ quarks $Q$ in the fundamental representation and $N_f$ anti-quarks $\tilde{Q}$ in the anti-fundamental representation) and the $SU(N_f-N_c) \equiv SU(\tilde{N}_c)$ magnetic theory ($N_f$ dual quarks $q$ and anti-quarks $\tilde{q}$ together with a meson field $M$) are supposed to describe the same physics in the infrared (IR). We assume that the equivalence persists sufficiently near the IR fixed point, and take $m \ll \Lambda~(\tilde{\Lambda})$ to justify this assumption, where $\Lambda$ ($\tilde{\Lambda}$) is the dynamical scale of the electric (magnetic) theory. 

The magnetic theory has the superpotential
\begin{align}
	W &= \frac{1}{\mu_{\rm m}} M^{ij} \tilde{q}_i q_j,\\
    \mu_{\rm m}^{N_f} &= \Lambda^{3N_c-N_f} \tilde{\Lambda}^{3\tilde{N}_c-N_f}\label{eq:matchingscale}
\end{align}
with $M^{ij}$ identified as the composite $ \tilde{Q}^i Q^j$ of the electric quarks and $\mu_{\rm m}$ the matching scale. The AMSB-perturbed version of this theory is calculable (under certain conditions) as we show in~\Cref{sec:Vacuummagnetic}.

The electric theory has no superpotential, and as we show in~\Cref{sec:RGEoriginalelectric} the AMSB-perturbed theory can be shown to run to a QCD-like phase in the IR. The low energy dynamics in this description is not calculable, but there is an alternative description available. Consistency of Seiberg duality implies that taking the dual of the magnetic theory again should return the original electric theory. This ``Twice-Dual" theory\footnote{We adopt the terminology ``Electric", ``Magnetic", and ``Twice-Dual" to distinguish between the different descriptions of the IR dynamics of the AMSB theory in the Conformal Window.} has the following superpotential
\begin{align}
 W=N^{ij}(Q_i\tilde{Q}_j-M_{ij}). 
\end{align}
The superfield $N^{ij}$ plays the role of a Lagrange multiplier. The equations of motion identify the meson $M$ with $Q_i\tilde{Q}_j$ and $N^{ij}=0$, which indeed reproduces the original electric theory. However, the presence of SUSY-breaking can modify this. For example, in the Intriligator-Seiberg-Shih model~\cite{Intriligator:2006dd}, the full rank of the meson matrix is supposed even if $N_f>N_c$ to  be consistent with the pure super Yang-Mills description after decoupling quarks. In our case, AMSB modifies the dynamics and gives $N$ a VEV which serves as a non-trivial modulus for the vacuum of the AMSB theory. Assuming that, for sufficiently small SUSY-breaking, the AMSB theory should respect Seiberg duality, we can analyze the IR of the Electric Theory by applying AMSB to the Twice-Dual.

For a superconformal theory, the conformal dimensions of chiral fields are determined completely\footnote{In general there can be other, non-anomalous $U(1)$ symmetries that make the choice of $U(1)_R$ charges ambiguous. The correct prescription is given by $a$-maximization~\cite{Intriligator:2003jj}. In this work, only the quarks are charged under the gauge group so the NSVZ $\beta$-function is sufficient to determine the fixed-point anomalous dimension of the quarks. That of the singlet (when present) can then be determined from the running of the Yukawa coupling. For this reason we do not need to invoke $a$-maximization.} by their $R$-charges, $D(\phi)=\frac{3}{2}R(\phi)$. On the other hand, the $U(1)_R$ symmetry in SUSY-QCD is determined by the anomaly-free condition and the charge conjugation invariance,
\begin{align}
	R(Q)&= R(\tilde{Q}) = \frac{N_f-N_c}{N_f}, \nonumber \\
	R(q) &= R(\tilde{q}) = \frac{N_f-\tilde{N}_c}{N_f}, \qquad
	R(M) = 2\frac{\tilde{N}_c}{N_f}\ .
\end{align} 
Given the ``AMSB convention" for the anomalous dimension, we have $\gamma = 1 - D(\phi)$, and therefore the K\"ahler potential receives the wave function renormalization
\begin{align}
	K = Z_\phi(\mu)\phi^* \phi = \left( \frac{\mu}{\Lambda} \right)^{2(1 - D(\phi))} \phi^* \phi.
	\label{eq:anomalous}
\end{align}
Here, $\mu$ is the renormalization scale, and $\Lambda$ is the energy scale where theory becomes nearly superconformal.

It is clear that AMSB effects asymptotically vanish toward an IR fixed point because couplings no longer run (see~\cref{eq:m2}),
\begin{align}
	m_{Q,q,M}^2(\mu) \rightarrow 0, &\qquad
	m_\lambda(\mu) \rightarrow 0.
\end{align}
However, the effects are relevant and change the IR dynamics unless
\begin{align}
	\frac{m_{Q,q,M}^2(\mu)}{\mu^2} \rightarrow 0, &\qquad
	\frac{m_\lambda(\mu)}{\mu} \rightarrow 0.
\end{align}
That is, if the AMSB effects scale sufficiently slowly as $\mu\to0$, they can produce different IR dynamics from the ordinary SUSY case. If not, the AMSB theory develops an ``emergent supersymmetry" in the IR. If and when the latter can happen is an ongoing area of research. This procedure and definition of relevance are standard, and for example can be seen in section 1.1 of~\cite{Rychkov_2017}.

 Unfortunately, we do not have computational tools to answer the question of relevance for the entire range of the conformal window. Instead, we look at Banks--Zaks (BZ) fixed points~\cite{Caswell:1974gg,Banks:1981nn} where the conformal dynamics can be studied using perturbation theory. This is possible at the upper edge or the lower edge of the conformal window as the IR fixed point couplings turn out to be perturbative, and we find that AMSB effects are relevant within the validity of such analysis.

\section{RGE Analysis around the Electric Banks--Zaks Fixed Point}\label{sec:RGEoriginalelectric}

\begin{figure}[ht]
    \centering
    \includegraphics[width=0.95\linewidth]{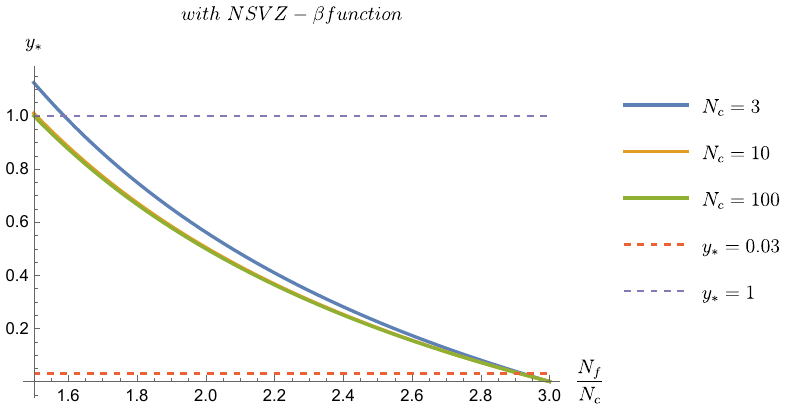}
    \caption{Plot of the gauge coupling $y_*=N_cg_*^2/(8\pi^2)$ at the fixed point as a function of $N_f/N_c$. }
    \label{fig:originalelecoupling}
\end{figure}

In this section, we show that the AMSB effects are relevant and modify the IR dynamics at the higher edge of the conformal window, using the electric BZ fixed point for $N_f = 3N_c/(1+\epsilon)$, $0 < \epsilon \ll 1$. We shall show that the squark mass-squared is positive, which would mean that the gauginos and squarks simply decouple and the theory reduces to the non-SUSY QCD at low energies, leaving no good handle on dynamics. \\
\ Following~\cite{Martin:1993zk}, running effects in the electric theory are described by the NSVZ $\beta$-function
\begin{align}
	\beta(g) &= \mu\frac{d}{d\mu}g^2 
	= -g^4 \frac{3N_c - N_f - N_f \gamma_Q}{8\pi^2 - N_c g^2}\ , \\
    \gamma_Q &= \frac{1}{8\pi^2} (2g^2 C_F),
\end{align}
with the quadratic casimir $C_F = \frac{N_c^2-1}{2N_c}$.  
Using 
$N_f =\frac{3N_c}{1+\epsilon},\ 
y \equiv \frac{N_c}{8\pi^2} g^2,$ 
the RGE is
\begin{align}
	\mu \frac{d}{d\mu} y 
    &= -3y^2 ( \epsilon - y), \\
	m_Q^2 
    &= \frac{3}{4} y^2 (\epsilon-y) m^2,
\end{align}
to the leading order in $\epsilon$ and large $N_c$. Note that since the electric theory is weakly coupled in the UV, $y < \epsilon$ as we flow to the IR fixed point. 
We show the coupling at the fixed point in~\cref{fig:originalelecoupling}.

\begin{figure}[ht]
\centering
\includegraphics[width=0.61\columnwidth]{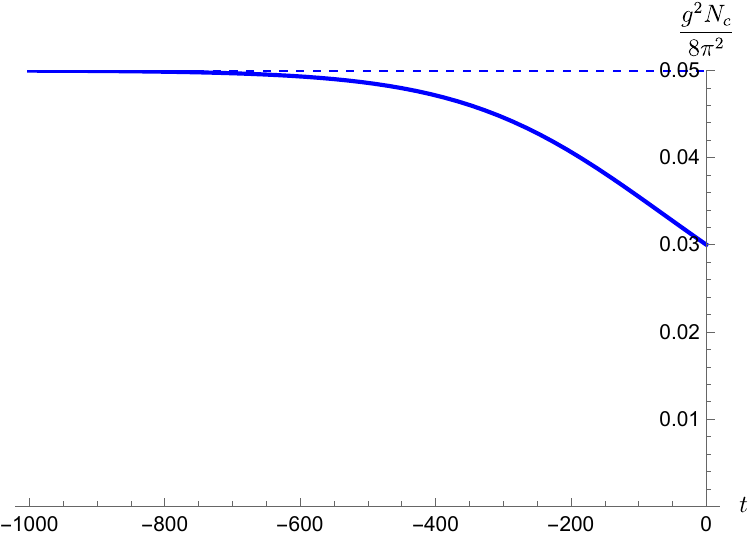}\\
\includegraphics[width=0.61\columnwidth]{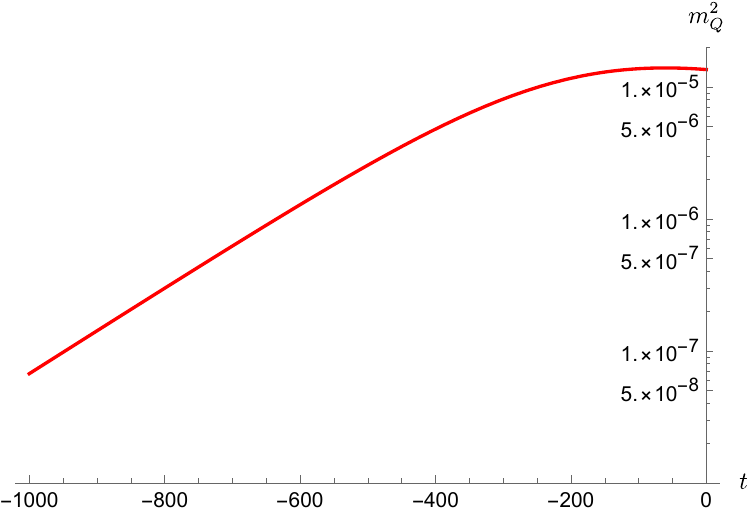}\\
\caption{Running of the electric gauge coupling (above) and the AMSB quark mass squared (below) in units with $m=1$ near the electric Banks--Zaks fixed point with $\epsilon=1/20$ and $N_c g^2(0)/8\pi^2 = 0.03$ and $N_c\gg 1$. The dashed line represents the infrared fixed point.
}
\label{fig:running}
\end{figure}

We can work out the approximate solution near the fixed point as
\begin{align}
	y(t) - \epsilon &=  (y(0)-\epsilon) e^{3\epsilon^2 t}, 
	\label{eq:y} \\
	m_Q^2 &= \frac{3}{4} \epsilon^2 (y(0)-\epsilon) e^{3\epsilon^2 t} m^2 .
	\label{eq:mQ2}
\end{align}
Here $t = \ln \mu \rightarrow -\infty$ defines the IR limit. An example of the solutions is shown in~\cref{fig:running}. AMSB is naively relevant so long as $3\epsilon^2 < 2$, or equivalently $N_f \gtrsim 1.65 N_c$. But this lower bound cannot be taken seriously because it occurs at $\epsilon \approx 0.8$, well beyond the point when the perturbative analysis breaks down.

Since $m_Q^2$ stays positive through all energy scales, and the only contribution to the scalar potential comes from the mass-squared term, it would naively appear that the potential minimum lies at the origin. However, the above analysis shows that for sufficiently small $\epsilon$, after a certain stage in the RG flow, $m_Q^2$ will exceed $\mu$ and then the squarks (and likewise gauginos) would decouple from the IR effective theory at some finite scale. Below that scale, the IR theory is that of non-SUSY QCD. The gauge coupling will begin to run to larger values and the theory will become strongly coupled. So analysis in terms of the electric description is inconclusive. \footnote{We remind that though we set out close to the perturbative regime of the Banks-Zaks fixed point at $N_f \approx 3N_c$ in the SUSY theory, this is not close to the Banks-Zaks fixed point in the non-SUSY theory, which typically appears near $N_f \approx 5.5N_c$. In other words, we would end up in the strongly coupled regime of the non-SUSY theory because of the threshold correction to the $\beta$-function.}

\section{RGE Analysis around the Magnetic Banks--Zaks Fixed Point}\label{sec:RGEmagnetic}

We again show that AMSB effects are relevant and modify the IR dynamics, in the case of magnetic BZ fixed points for $N_f = \frac{3\tilde{N}_c}{1+\tilde\epsilon}$, $0 < \tilde\epsilon \ll 1$ with $\tilde{N}_c=N_f-N_c$. This is close to the lower edge of the conformal window $N_f \approx \frac{3}{2}N_c$.

Following~\cite{deGouvea:1998ft}, the magnetic RGEs are given by
\begin{align}
	\gamma_q &= \frac{1}{8\pi^2} (2g^2 C_F - \lambda^2 N_f), \\
	\gamma_M &= -\frac{1}{8\pi^2} \lambda^2 \tilde{N}_c, \\
	\beta_g &= -g^4 \frac{3\tilde{N}_c - N_f - N_f \gamma_q}{8\pi^2 - \tilde{N}_c g^2}, \\
	\beta_\lambda &= -\lambda^2 (2\gamma_q + \gamma_M),
\end{align}
with $C_F =\frac{\tilde{N}_c^2-1}{2\tilde{N}_c}$, and the squark and meson mass-squared are given by
\begin{align}
	m_q^2 =& -\frac{1}{4}\dot{\gamma}_q m^2 = \frac{1}{32\pi^2}\left(
	-2\beta_g C_F+ \beta_\lambda N_f
	\right)m^2,\\
	m_M^2 =& -\frac{1}{4}\dot{\gamma}_Mm^2 = \frac{1}{32\pi^2}\beta_\lambda \tilde{N}_c.
\end{align}
Using
$x \equiv \frac{\tilde{N}_c}{8\pi^2} \lambda^2,\ 
y \equiv \frac{\tilde{N}_c}{8\pi^2} g^2$, and $N_f=3\tilde{N}_c/(1+\tilde{\epsilon})$,
the RGE is
\begin{align}
     \frac{dx}{d\log\mu} 
&= x\left(7x-2y\right),\label{eq:magRGE1}\\
 \frac{dy}{d\log\mu}  
 &= -3y^2(3x-y+\tilde{\epsilon})\label{eq:magRGE2},
\end{align}
where we took $\tilde{\epsilon}\ll 1$ and large $\tilde{N}_c$ to make perturbativity better.

\begin{figure}[ht]
    \centering
    \includegraphics[width=0.9\linewidth]{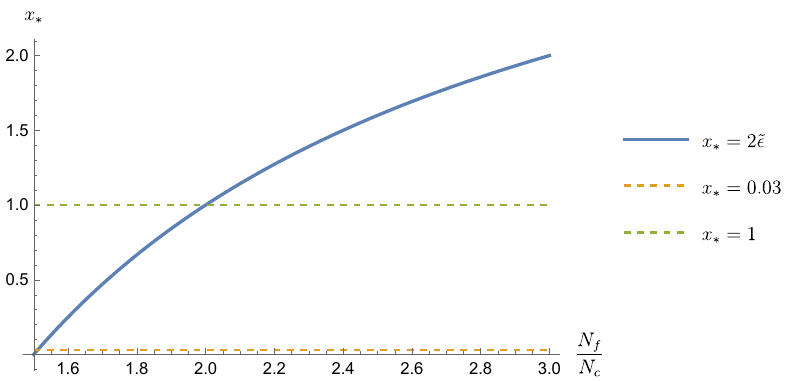}
    \caption{The value of the coupling $x_*=\tilde{N}_c\lambda^2/8\pi^2$ at the fixed point as a function of $N_f/N_c$, obtained using the 1-loop RGE.}
    \label{fig:maglambdacoupling}
\end{figure}

\begin{figure}
    \centering
    \includegraphics[width=0.9\linewidth]{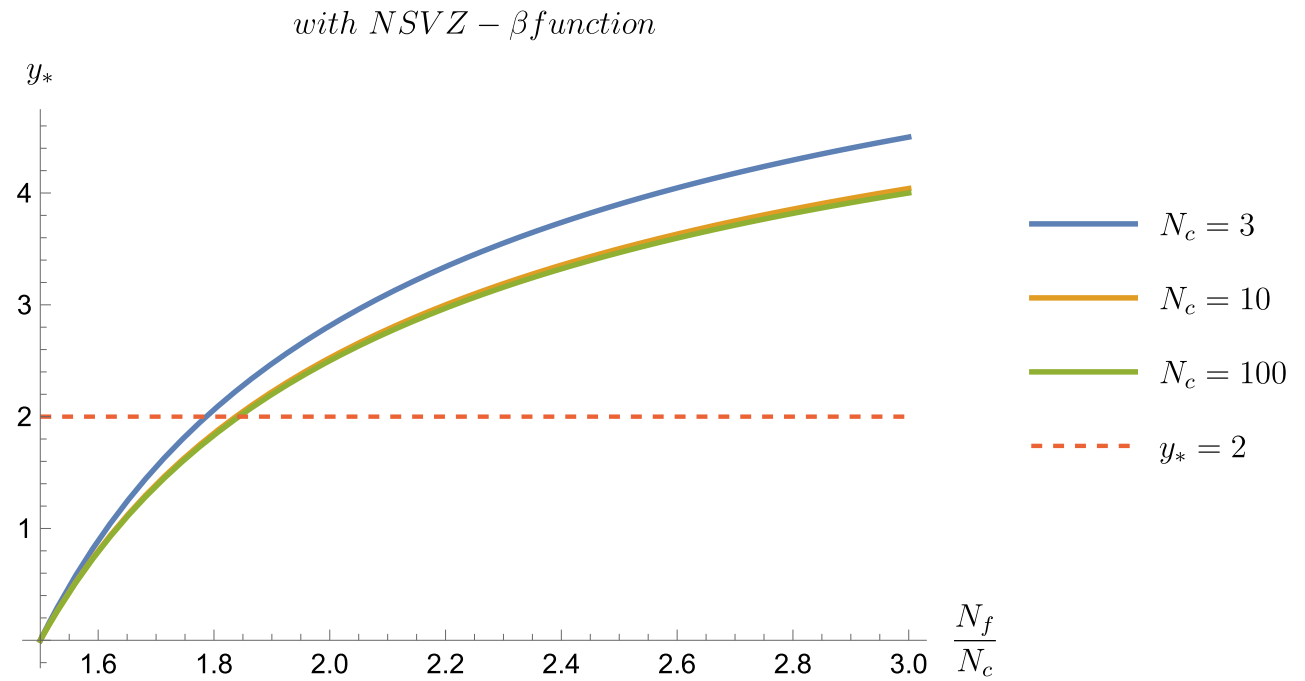}
    \caption{The value of the coupling $y_*=\tilde{N}_cg^2/8\pi^2$ at the fixed point as a function of $N_f/N_c$, obtained using the 1-loop RGE.}
    \label{fig:maggaugecouppling}
\end{figure}

The fixed fixed point of the RGE in~\cref{eq:magRGE1,eq:magRGE2} is
\begin{align}
	(x_*,y_*) =& \left(2\tilde{\epsilon},7\tilde{\epsilon}\right). 
	\label{eq:exactBZ}
\end{align}
We show the values of these couplings at the fixed point as a function of $\frac{N_f}{N_c}$ in~\cref{fig:maglambdacoupling,fig:maggaugecouppling}.
Around this fixed point, we can expand the $\beta$-function
\begin{align}
\frac{d}{d\log\mu}\begin{pmatrix} x- x_* \\ y-y_* \end{pmatrix} 
&= \beta_D  \begin{pmatrix}  x-x_*\\ y-y_*  \end{pmatrix} \\
\beta_D&\equiv
\begin{pmatrix} 14\tilde{\epsilon} & -4\tilde{\epsilon}  \\ -441\tilde{\epsilon}^2   
& 147\tilde{\epsilon}^2  \end{pmatrix}
\end{align}

The eigenvalues $(\lambda_1,\lambda_2)$ of $\beta_D$ are
\begin{align}
\lambda_1 &\simeq 14\tilde{\epsilon}+114\tilde{\epsilon}^2+\frac{126}{\tilde{N}_c^2} \tilde{\epsilon}^2+\frac{126}{\tilde{N}_c}\tilde{\epsilon}^2+126\tilde{N}_c\tilde{\epsilon}^2 \\
\lambda_2&\simeq 21\tilde{\epsilon}^2 +\frac{21}{\tilde{N}_c^2} \tilde{\epsilon}^2-\frac{126}{\tilde{N}_c} \tilde{\epsilon}^2-126\tilde{N}_c\tilde{\epsilon}^2
\end{align}

The eigenvalues of this matrix are $\mathcal{O}(\tilde{\epsilon})$ and $\mathcal{O}(\tilde{\epsilon}^2)$. Notice that these exponentials are slower than $\mu^2 = e^{2t}$ whenever $\lambda_1, \lambda_2 < 2$. Therefore the AMSB effects are relevant for sufficiently small $\tilde{\epsilon}$. As $\tilde{\epsilon}$ increases, we can no longer trust the perturbative analysis and cannot make a conclusion.

The corresponding trajectories are
\begin{align}
    -\frac{7}{2}\left(1-\frac{3}{2}\tilde{\epsilon}\right)\delta x + \delta y\  \text{for}\  \lambda_1\simeq14\tilde{\epsilon}, \label{eq:faster}\\
    \frac{63}{2}\tilde{\epsilon}\delta x + \delta y\ \text{for}\  \lambda_2\simeq21\tilde{\epsilon}^2 \label{eq:slower},  
\end{align}
where we have defined $\delta x = x- x_*$ and $\delta y = y- y_*$. Since $21\tilde{\epsilon}^2 \ll 14\tilde{\epsilon}$ the combination in~\cref{eq:faster} quickly runs to zero, leaving the remainder of the flow with the slower eigenvalue to occur along the line: 
\begin{align}
    \delta x =\frac{2}{7}& \delta y \frac{1}{1-\frac{3}{2}\tilde{\epsilon}} \approx \frac{2}{7}\left(1+\frac{3\tilde{\epsilon}}{2}\right)\delta y
\end{align}

Using this result to express $\beta (x)$, $\beta(y)$ in terms of $\delta y$ and $\tilde{\epsilon}$, we get the RG flow
\begin{align}
    \beta (y) &= 21\tilde{\epsilon}^{2}\delta y \\
    \beta(x) &= 6\tilde{\epsilon}^{2}\delta y \approx 21\tilde{\epsilon}^{2}\delta x 
\end{align}
yielding 
\begin{align}
  \delta x\sim \delta y \sim \mu^{21\tilde{\epsilon}^{2}} 
\end{align}
Then the meson and dual squark masses are
\begin{align}
    m_{M}^{2} &= \frac{3}{2}\tilde{\epsilon}^{2} \delta y m^{2} \\
    m_{q}^{2} &= -\frac{3}{4}\tilde{\epsilon}^{2}\delta y m^{2} \label{eq:massesE}
\end{align}
Notice that if we approach the fixed point from below in coupling space, then $\delta y < 0$ and hence $m_M^{2} < 0$, $m_q^{2} > 0$. 

\begin{figure}[ht]
\centering
\includegraphics[width=0.84\linewidth]{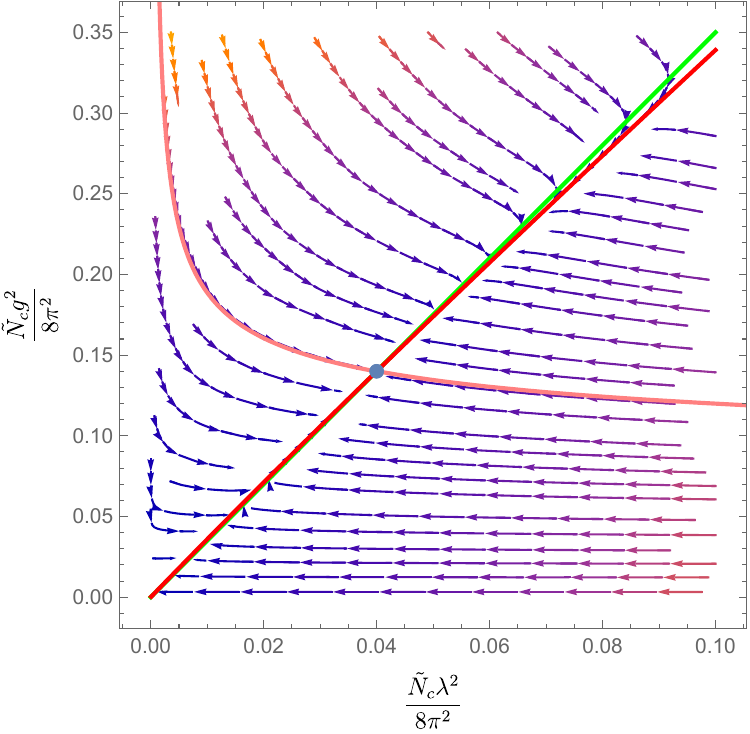}
\caption{Two-dimensional RGE flow of coupling constants near the magnetic Banks--Zaks fixed point (blue dot) with $\tilde{\epsilon}=1/50$. $m_{\tilde{q}}^2 > 0$ below the red line, while $m_M^2 > 0$ below the green line. The pink curve separates regions of initial conditions, with those below the curve leading to the $\chi$SB minimum.
 }
\label{fig:2DRG}
\end{figure}

\begin{figure}[ht]
\centering
\includegraphics[width=0.69\linewidth]{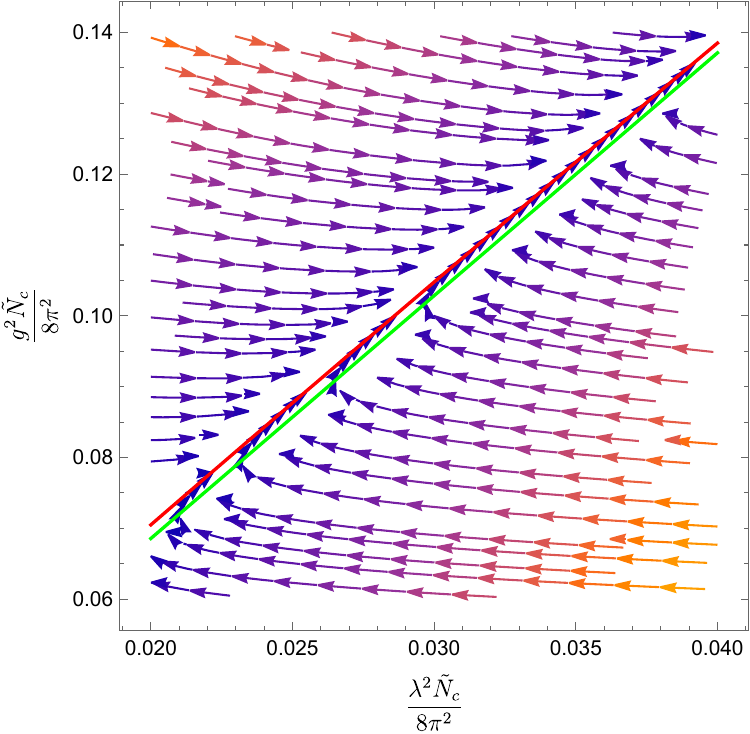}
\caption{Zoomed up version of ~\Cref{fig:2DRG} to the left of the fixed point, with $\tilde{\epsilon}=1/38$ (just to aid visibility). It can be seen that the flow towards the fixed point falls into the region where $m_M^2<0$ and $m_Q^2>0$, and first converges on the $m_{q}^{2}=0 $ (red) line. There is also a clear indication the RGE flows follow  the fast $\mathcal{O}(\tilde{\epsilon}^2)$ eigen-trajectory and converge toward the fixed point along the slower eigen-trajectory.}
\label{fig:2DRGzoomup}
\end{figure}

\begin{figure}[ht]
\centering
\begin{subfigure}
 \centering
 \includegraphics[width=0.49\linewidth]{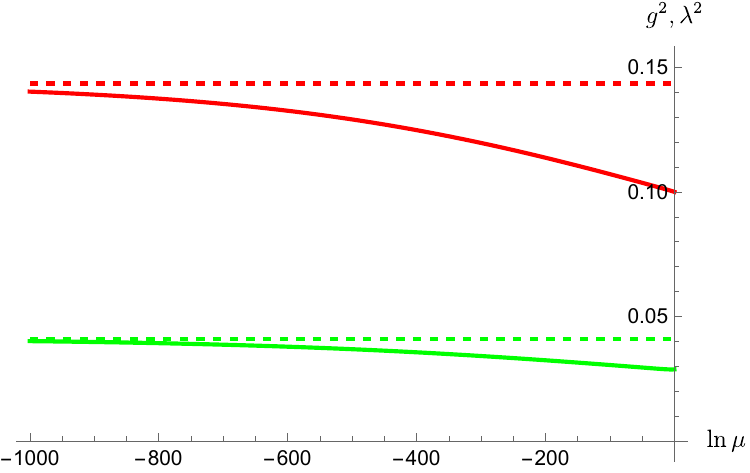}
\end{subfigure}
\begin{subfigure}
 \centering
 \includegraphics[width=0.49\linewidth]{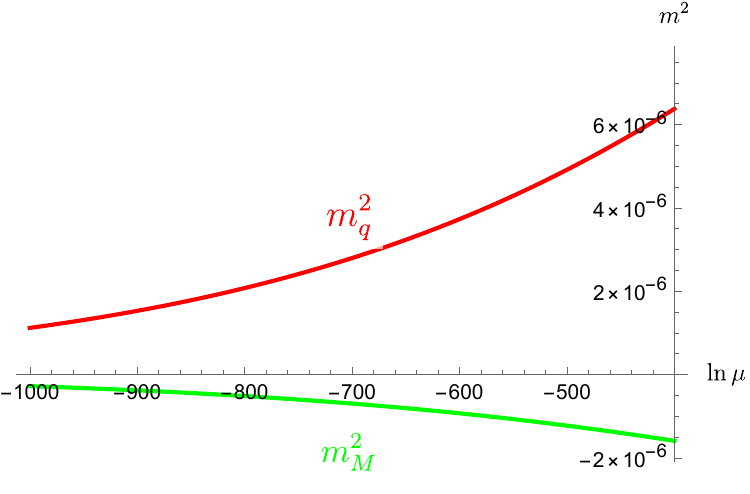}
\end{subfigure}
\caption{Left: Running of $g^2$ (red) and $\lambda^2$ (green),  
with the initial condition $g^2 = 0.1$, $\lambda^2 = \frac{2}{7} g^2 + 0.00015$, while the dashed lines show their infrared fixed point values. We took $N_c = 100$, $N_f=151$, $\tilde{N}_c=51$, and hence $\tilde{\epsilon} \approx 0.0132$. 
Right: Corresponding running of $m_q^2$ (red) and $m_M^2$ (green) in units with $m=1$. 
The slow exponent $e^{21\tilde{\epsilon}^2 t}$ is apparent and both AMSB effects are a relevant perturbation. }
\label{fig:runmass}
\end{figure}

A numerical plot of the RG flow is shown in~\cref{fig:2DRG,fig:2DRGzoomup}, superposed with curves of $m_q^2 = 0$ (red) and $m_M^2 = 0$ (green). Notice that the flows appear to converge on the $m_{q}^2 = 0$ line first, providing an indication of the finite meson expectation value in the ground state. A separate illustration of the running couplings and its consequences for the squark and meson masses -- in the regime where $m_q^2>0$ -- is shown in~\cref{fig:runmass}. It is reasonable to expect that the initial conditions are near the UV fixed point $g=\lambda=0$. For numerical plots, we took values already close to the IR fixed point to keep the amount of running manageable.

It might appear that the dynamics is ambiguous because depending on the initial condition of coupling constants, $m_q^2$ and $m_M^2$ are found to have either sign. However, in requiring that baryon direction \footnote{HM thanks Nathaniel Craig for pointing out a potential instability along this direction. HM also thanks Csaba Cs\'aki and Ofri Telem for pointing out that there is no baryon direction for magnetic $SO(N_c)$ or $Sp(N_c)$ theories. {This observation amplifies the confidence that such an instability would not occur for $SU(N_c)$ theories either.}} {does not runaway to infinity} in a way that is consistent with the electric theory, we have to assume that the initial conditions lie in the region of the parameter space where $m_q^2 \geq 0$. Otherwise the dual quark would roll down the potential to a minimum of $O\left(-\left(\frac{m}{4\pi}\right)^4\right)$ or possibly even $O\left(-\frac{m^2\tilde{\Lambda}^2}{(4\pi)^4}\right)$, and break the chiral symmetry $SU(N_f)_{q} \times SU(N_f)_{\tilde{q}} \times U(1)_B$ to $SU(N_f-\tilde{N}_c)_{q} \times SU(N_f)_{\tilde{q}}$.

It is also not too surprising that the duality imposes a constraint on the magnetic theory, given that the electric theory has a single parameter $g$, while the magnetic theory two $g$ and $\lambda$. For them to describe the same physics, there must be a one-parameter constraint on the $(g,\lambda)$ parameter space. In the asymptotic IR limit, both theories reach the IR fixed point and there are no parameters. Slightly away from the IR fixed point, the magnetic theory should follow a specific RG trajectory on the $(g,\lambda)$ plane. In the supersymmetric limit, we did not have tools to pick a trajectory. Together with the AMSB effects, now we can at least choose a half of the parameter space. Existing studies claim to pick out a particular trajectory from consistency with the duality, see~\cite{Oehme:1999fd}. While we do not assume such a trajectory, our results above agree with that trajectory at sufficiently low energies, regardless of initial conditions.\\ 

\section{RGE Analysis around the Electric Banks-Zaks Fixed Point in the Twice-Dual Description}\label{sec:RGEelectricTwiceDual}

In this section, we analyze the RGE for the electric theory after taking dual twice. Since the presence of mass term explicitly breaks the conformal symmetry, it is sufficient to study the gauge coupling and Yukawa coupling. (The mass will run to zero in the IR, as can be seen by the fact that it is only renormalized by the wavefunciton of $N$.) 
We analyze the following 1 loop level RGE~\cite{Martin:1993zk},

\begin{align}
\gamma_Q &= \gamma_{\tilde{Q}}
= \frac{1}{8\pi^{2}}\left(2g^2\frac{N_c^2-1}{2N_c}-N_f|Y_{NQ\tilde{Q}}|^2\right) \\
\gamma_N &= -\frac{N_c}{8\pi^{2}}|Y_{NQ\tilde{Q}}|^2 \\
\beta_{g} &= \frac{d}{dt} g^2 
=-g^4 \frac{3N_c-N_f-\frac{N_f}{2}(\gamma_Q+\gamma_{\tilde{Q}}) }{8\pi^2-N_cg^2} \nonumber\\
\beta_Y &= \frac{d}{dt}|Y_{NQ\tilde{Q}}|^2 =-|Y_{NQ\tilde{Q}}|^2 \left(2\gamma_Q+\gamma_N\right)
\end{align}
using $x\equiv \frac{N_c|Y_{NQ\tilde{Q}}|^2}{8\pi^2},\ y\equiv \frac{N_c g^2}{8\pi^2}$, and taking $N_c$ large:
\begin{align}
\frac{d}{dt} x &= - x\left(
2\left(y-\frac{N_f}{N_c}x\right)
-x\right)\\
\frac{d}{dt}y&= -\frac{y^2}{1-y}\left(3-\frac{N_f}{N_c}\left(1+y-\frac{N_f}{N_c}x\right)\right)
\end{align}
The fixed point $(x_*,y_*)$ is determined by $\frac{dx}{dt}=0, \frac{dy}{dt}=0$. Taking $N_f =\frac{3N_c}{1+\epsilon}$ we have to leading order
\begin{align}
(x_*,y_*) =& (2\epsilon,7\epsilon)
\end{align} 
Consider linear deviations around this fixed point $x=x_*+\delta x,\ y=y_*+\delta y$. The linearized RGE is given by
\begin{align}
    \dod{}{t}\begin{pmatrix}
        \delta x \\ \delta y
    \end{pmatrix}
    =& \begin{pmatrix}
        14\epsilon & -4\epsilon \\
         -441\epsilon^2 & 147\epsilon^2
    \end{pmatrix}\begin{pmatrix}
        \delta x \\ \delta y
    \end{pmatrix}
\end{align}
The eigenvalues of this matrix are $14\epsilon$ and $21\epsilon^2$. It is not surprising that this looks just like the result in the magnetic theory: as the fixed point is approached the superpotential mass term runs to zero and we are left with the same low energy theory as the magnetic case, albeit with $N_c,\epsilon$ and different labels for the fields. As in the magnetic case, as one flows toward the fixed point the larger exponent $14\epsilon$ will decay rapidly, leaving the final approach along the direction corresponding to $21\epsilon^2$. The dynamics is again ambiguous because depending on the initial condition of coupling constants, $m_Q^2$ and $m_N^2$ are found to have either sign. However, if $m_Q^2 < 0$, squark VEVs $Q = v\delta_{ij}$ (for $i, j = 1, \cdots, N_c$) will turn on. In requiring that this baryon direction does not runaway to $\mathcal{O}(\Lambda)$ field values \footnote{We know that $m_Q^2 > 0$ in the UV description, so the ``runaway" stops at $\mathcal{O}(\Lambda)$.}, we have to assume that the initial conditions lie in the region of the parameter space where $m_Q^2 > 0$ and $m_N^2 < 0$ (the corresponding flows approach the fixed point from below in coupling space). This assumption is also backed by our RGE analysis in the electric theory (see~\Cref{sec:RGEoriginalelectric}), where we unambiguously show that $m_Q^2 > 0$ in the viscinity of the electric Banks-Zaks fixed point. \\

\begin{figure}[ht]
    \centering
\includegraphics[width=0.813\linewidth]{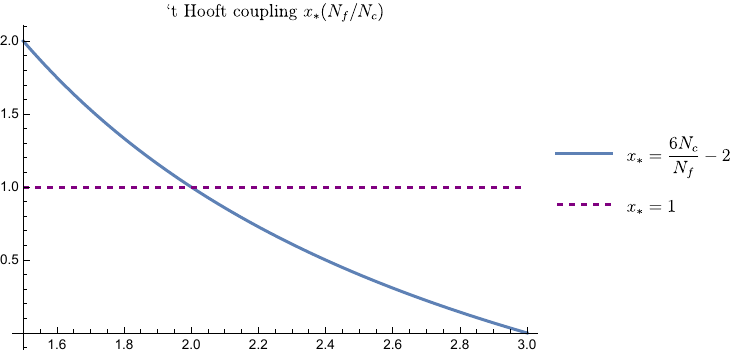}
    \caption{The critical value of `t Hooft coupling $x_*=\frac{N_c\abs{Y_{NQ\bar{Q}}}^2_*}{8\pi^2}$ as a function of $\frac{N_f}{N_c}$ at fixed point for the twice-dual theory, obtained via the 1-loop anomalous dimensions without a large $N$ limit.}
    \label{fig:TWDYukawa}
\end{figure}

\begin{figure}[ht]
    \centering
\includegraphics[width=0.813\linewidth]{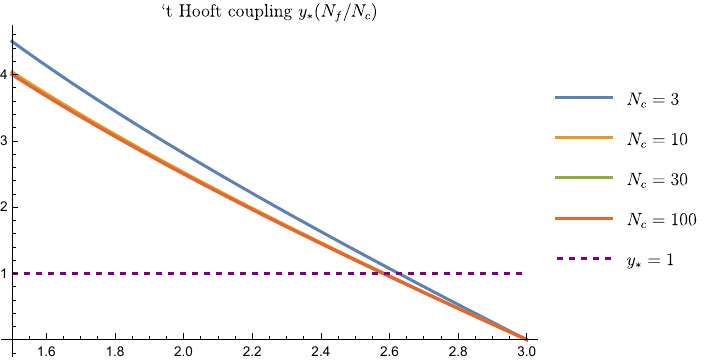}
    \caption{The critical value of `t Hooft coupling $y_*=\frac{N_c g_*^2}{8\pi^2}$ as a function of $\frac{N_f}{N_c}$ at fixed point for the twic-dual theory, obtained via the 1-loop anomalous dimensions without a large $N$ limit.}
    \label{fig:Beta2}
\end{figure}

We show the critical value of the `t Hooft couplings in \cref{fig:TWDYukawa,fig:Beta2}, calculated from the NSVZ $\beta$-function with 1-loop anomalous dimensions, as a function of $\frac{N_f}{N_c}$.

\newpage
\section{Chiral Symmetry Breaking in the Twice Dual Electric Theory}\label{sec:VacuumelectricTwicedual}

\begin{figure}[ht]
    \centering
    \includegraphics[width=0.8\linewidth]{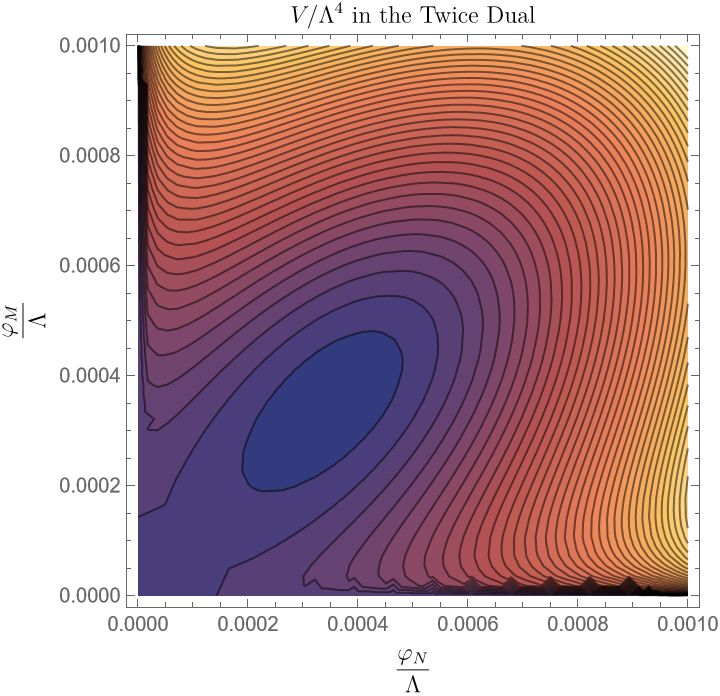}
    \caption{The plot of the potential indicating the minimum at $\varphi_M=\varphi_N$ in~\cref{eq:soleleTwD}. We took $N_c=5,N_f=11$, $m=0.1\Lambda,Y=1$, and $\mu=\sqrt{\varphi_M\varphi_N}$ as an example. Note that this is the full potential \cref{eq:fullpotTWDq} (with $c_M=c_N=c_\gamma=1$) without any truncation, large $N$ or small $\epsilon$ approximations. Different choices of $N_f,N_c$ give analogous results. Different choices of $\mu$ slightly deform the contours, but keep the minimum at $\varphi_M=\varphi_N$.}
    \label{fig:PotentialTWD}
\end{figure}

Given that the AMSB effects are relevant in the vicinity of the upper edge of the conformal window, we shall now show that in the IR limit of the twice-dual theory, a chiral symmetry broken vacuum can be realized. We parametrize $N_f=\frac{3N_c}{1+\epsilon},\ (\epsilon\ll 1)$. Before starting calculations, we comment that the solution of $M,N$ is close to zero $\left(\frac{M}{\Lambda},\frac{N}{\Lambda}\ll 1\right)$. At the edge of the window $N_f=3N_c$ $(\epsilon=0)$, where the superconformal symmetry is restored, the solution is $M=N=0$. Assuming $m_Q^2 > 0$ and after decoupling the squarks, we start from the following superpotential
 \begin{align}\label{eq:SuperPotTWDEFT}
W
&= N_c\Lambda^3 \left( \frac{Y \phi_N}{\Lambda} \right)^{\frac{N_f}{N_c}}-N_fY\Lambda \phi_N\phi_M,  
 \end{align}
where meson $M,N$ is taken to be a diagonal form to realize the chiral symmetry broken vacuum $M=\phi_M\delta_{ij},N=\phi_N\delta_{ij}$. Following the method to obtain~\cref{eq:AMSBpotential} in~\Cref{sec:AMSB}, we can calculate the potential. The calculation is done out in~\Cref{sec:genericmu}. We take canonically normalized fields $\phi_M\rightarrow\frac{\phi_M}{\sqrt{c_MN_f}},\ \phi_N\rightarrow\frac{\phi_N}{\sqrt{c_NN_f{Z}_N}}$. After using the rotation of $U(1)_R$ to rotate the phase of $m$, the potential is to leading order
\begin{align}\label{eq:TWDmainpotential}
\frac{V}{\Lambda^4} \simeq
&\frac{Y^2}{c_Nc_M}\left(\frac{\mu}{\Lambda}\right)^{2\epsilon}\left(\frac{\phi_M}{\Lambda}\right)^2
+\frac{Y^2}{c_Mc_N}\left(\frac{\mu}{\Lambda} \right)^{2\epsilon}\left( \frac{\phi_N}{\Lambda}\right)^2 \nonumber\\
&-\frac{2Y}{\sqrt{c_Mc_N}}\frac{m}{\Lambda}\left( \frac{\mu}{\Lambda}\right)^{\epsilon} \frac{\phi_M}{\Lambda}\frac{\phi_N}{\Lambda}
\end{align}
This is symmetric in $\phi_M$ and $\phi_N$ and we expect that the minimum is along $\mu=\phi_N=\phi_M\equiv\phi$. This is justified via numerical results and analytical results in~\Cref{sec:genericmu,sec:TDWODecoupleQ}. The potential is then
\begin{align}
\frac{V}{\Lambda^4} \simeq
&\frac{2Y^2}{c_Nc_M}\left(\frac{\phi}{\Lambda}\right)^{2+2\epsilon}
-\frac{2Y}{\sqrt{c_Mc_N}}\frac{m}{\Lambda}\left(\frac{\phi}{\Lambda}\right)^{2+\epsilon}
\end{align}
The minimum of this potential is
\begin{align}\label{eq:soleleTwD}
\frac{\phi}{\Lambda}&=
\left(\frac{\sqrt{c_Mc_N}}{Y}\frac{m}{\Lambda}\right)^{\frac{1}{\epsilon}}.   
\end{align}
With this solution, we can obtain the scaling of the vacuum energy $V/\Lambda^4$ in terms of $m/\Lambda$
\begin{align}\label{eq:scalingeleTwD}
    \frac{V}{\Lambda^4}\propto \left(\frac{m}{\Lambda}\right)^{2+\frac{2}{\epsilon}}.
\end{align}
We show the potential in~\cref{fig:PotentialTWD}. This result numerically supports our simple analysis. The calculation process with equation of motion for a general choice of energy scale is in~\Cref{sec:genericmu}. Another analysis including quarks $Q,\tilde{Q}$ shows the same solution, which is in~\Cref{sec:TDWODecoupleQ}. The fact that the same result can be found via multiple different approaches supports the conclusion of $\chi$SB near the upper edge of the window (when $m_Q^2>0$).

\section{Chiral Symmetry Breaking in the Magnetic Theory}\label{sec:Vacuummagnetic}

We argued that the AMSB effects are also relevant at the lower edge of the conformal window, which deflects the RGE flow away from the superconformal dynamics, and $m_q^2 > 0$ so that there is no baryonic instability with a vacuum energy of {deeper than} $-O\left(\left(\frac{m}{4\pi}\right)^4\right)$. Given our observations, we would again like to understand what the IR limit of the theory is. We show that the ground state appears with a finite meson expectation value.

Written in terms of dimension 1 fields, the superpotential of the dual theory is
\begin{align}
    W =& \lambda M \tilde{q}q
\end{align}
When the meson VEV is close to origin $M_m\sim 0$, regularity implies that the Kahler potential should taken to be
\begin{align}
K\simeq c_m Z_m(\mu) M^\dag M
\end{align}
where
\begin{align}\label{eq:anomalous2}
Z_m(\mu)=\left(\frac{\mu}{\tilde{\Lambda}}\right)^{\gamma_M}
\end{align}
is the wave function renormalization. 

After integrating out the dual squarks, the low-energy theory is just pure SYM and develops a gaugino condensate with the non-perturbative superpotential 
\begin{align}
W_{\text{eff}} &= \tilde{N}_c \left( \tilde{\Lambda}^{ 3\tilde{N}_c-N_f } \det(\lambda M) \right)^{\frac{1}{\tilde{N}_c}}
= \tilde{N}_c \left( \tilde{\Lambda}^{ 3\tilde{N}_c-N_f } \lambda^{N_f} \phi^{N_f} \right)^{\frac{1}{\tilde{N}_c}},
	\label{eq:W}
\end{align}
where $M_{ij} = \phi \delta_{ij}$. The scalar potential including loop-level AMSB effects is

\begin{align}
  V&=-\frac{N_f}{4}c_mZ_m\dot{\gamma}_Mm^2\phi^2+
m\left(1-\frac{\gamma_M}{2} \right)\left( \phi\frac{\partial W}{\partial\phi} +\phi^*\frac{\partial W^*}{\partial \phi^*}\right)\nonumber\\
&-3mW-3mW^*
+\frac{1}{N_fc_mZ_m} \left| \frac{\partial W}{\partial \phi} \right|^2.
 \end{align}

Note that AMSB part of this potential vanishes in the strict conformal limit. We will consider deviations from the conformal fixed point, parametrized by
\begin{align}
\gamma_M =-2\frac{3\tilde{N}_c-N_f}{N_f} + c_\gamma \left(\frac{\mu}{\tilde{\Lambda}}\right)^\alpha
\end{align}
We switch to the canonically normalized field $\tilde{\phi}= \phi \sqrt{c_mZ_m N_f}$ and assume $\tilde{\phi}\in\mathbb{R}$. For self-consistency, the potential is evaluated at $\mu \sim \tilde\phi$ as in the Coleman-Weinberg potential\ ~\cite{Coleman:1973jx}. The scalar potential is then
\begin{align}\label{eq:Vmag}
\frac{V}{\tilde{\Lambda}^4}=& N_f^2\left(\frac{\lambda^2}{c_m N_f}\right)^{\frac{N_f}{\tilde{N}_c}}  \left(\frac{\tilde{\phi}}{\tilde{\Lambda}}\right)^{4}
    - \frac{1}{4}c_\gamma \alpha \left(\frac{m}{\tilde{\Lambda}}\right)^2\left(\frac{\tilde{\phi}}{\tilde\Lambda}\right)^{2+\alpha}
    \nonumber\\
    &-c_\gamma N_f \left(\frac{\lambda}{\sqrt{c_m N_f}}\right)^{\frac{N_f}{\tilde{N}_c}} \frac{m}{\tilde{\Lambda}}\left(\frac{\tilde{\phi}}{\tilde{\Lambda}}\right)^{3+\alpha}
\end{align}
For $c_\gamma >0$ the minimum of this potential will be at $\tilde{\phi}\neq 0$. This implies that the RG trajectory must be such that $m_M^2 < 0$ near the fixed point, which is a non-trivial restriction on the space of bare couplings. This choice is, however, necessary for continuity between the results at the upper edge of the window in~\Cref{sec:VacuumelectricTwicedual} and results in the Free Magnetic phase \cite{Csaki:2022cyg}.

If we minimize the most relevant part of the potential, which for $\tilde{\phi} \sim m$ and $\alpha\ll 1$ includes only the first two terms in the RHS of~\cref{eq:Vmag}, we find
\begin{align}\label{eq:solmag}
    \frac{\tilde\phi}{\tilde{\Lambda}} =& \left(\frac{c_\gamma c_M^{\frac{N_f}{\tilde{N}_c}} \alpha(2+\alpha) m^2}{16 \lambda^{\frac{2N_f}{\tilde{N}_c}} N_f^{\frac{2\tilde{N}_c-N_f}{\tilde{N}_c}} \tilde{\Lambda}^2}\right)^{\frac{1}{2-\alpha}} \propto \left(\frac{m}{\tilde{\Lambda}}\right)^{\frac{2}{2-\alpha}}
\end{align}
Note that this minimum is indeed of order $m$ for $\alpha\ll 1$, justifying the minimization procedure a posteriori. With the solution~\cref{eq:solmag} the vacuum energy scales as
\begin{align}\label{eq:magalpha}
 \frac{V}{\Lambda^4}\propto \left(\frac{m}{\Lambda}\right)^{\frac{8}{2-\alpha}}   
\end{align}
Note that as $\alpha\to 0$ the minimum approaches the origin and the vacuum energy vanishes. We know from~\Cref{sec:RGEmagnetic} that for $\tilde{\epsilon}\ll 1$, we have $\alpha\sim 21\tilde\epsilon^2$, so the $\alpha=0$ point corresponds to the lower edge of the window $N_f=\frac{3N_c}{2}$. This indicates that the superconformal symmetry is restored at exactly $\frac{3N_c}{2}$ and that AMSB alone is not suitable for analyzing the non-SUSY limit. This agrees with the analysis starting from $N_f=\frac{3N_c}{2}$, where the superpotential is classically conformal and the AMSB-perturbed theory does not produce a $\chi$SB minimum (at least not for perturbative field values).

\section{A Comprehensive Picture of ASQCD}\label{sec:allASQCD}
In this section we summarize the conclusions made, in this paper and in previous works \cite{Csaki:2022cyg,Murayama:2021xfj}, from applying AMSB to the question of chiral symmetry breaking in QCD for various ranges of flavors $N_f$ and colors $N_c$.

For $N_f<N_c$ the SQCD theory is in the ADS phase. Here the AMSB-perturbed theory has a calculable $\chi$SB minimum that is stable. The massless particle spectrum consists of the Nambu-Goldstone bosons of the broken chiral symmetry and there is no sign of a phase transition as the SUSY breaking $m$ is increased.

For $N_f=N_c$ the SQCD theory is in the Quantum-Modified phase. Here the low energy dynamics is genuinely strongly coupled, such that higher-order terms in the K\"ahler potential cannot be ignored. A candidate $\chi$SB minimum for the AMSB-perturbed theory has been identified, but whether or not there are baryonic runaways at this point is subject to the ratio of unknown order-one coefficients in the K\"ahler potential.

For $N_f=N_c+1$ the SQCD theory is s-confining. Excluding the special case of $N_c=2$, the ASQCD theory has been shown to have a stable $\chi$SB minimum near the origin of moduli space. For $N_c=2$, the SQCD theory has a superpotential that is classically conformal invariant and AMSB does not appear to be sufficient to analyze the non-SUSY limit \cite{deLima:2023ebw}. 

For $N_c+2\leq N_f<\frac{3N_c}{2}$ the SQCD theory is in the Free Magnetic phase. It has been shown that for the majority of this phase, $N_f\lesssim 1.43 N_c$, the ASQCD theory has a stable $\chi$SB minimum without runaways. For $1.43N_c\lesssim N_f <\frac{3N_c}{2}$ the dual squark mass-squared turns negative and baryonic runaways a minimum with $q,\bar{q}\sim \Lambda$ are expected. A local $\chi$SB minimum is still present, however, and may be continuously connected to the non-SUSY limit. 

It should be noted that at exactly $N_f=\frac{3N_c}{2}$, the dynamically generated superpotential enjoys classical conformal invariance. This is known to obstruct the usefulness of AMSB \cite{deLima:2023ebw,Bai:2021tgl}, and so this particular point remains an open problem. One may, however, speculate about chiral symmetry breaking in the non-SUSY theory based on the results immediately above and slightly below in $N_f$.

Now we turn our attention to the conformal window, which has been the subject of this paper. At the lower edge of the window, $\frac{3N_c}{2}\lesssim N_f \ll 3N_c$, using the magnetic description, we have demonstrated that AMSB is relevant in the IR and deflects the RG flow toward a $\chi$SB minimum, subject to an assumption on the initial conditions of the RG flow (see~\Cref{sec:Vacuummagnetic}). We do the same at the top of the conformal window, $\frac{3N_c}{2} \ll N_f \lesssim 3N_c$, using the Twice-dual theory. Again depending on the initial conditions, the squarks get a positive or negative mass-squared from AMSB. However, in the latter case, the squark VEVs inevitably turn on, leading to a $\chi$SB minimum ($SU(N_f) \times SU(N_f) \rightarrow SU(N_c) \times SU(N_{f}-N_{c}) \times SU(N_f)$) at $\mathcal{O}(\Lambda)$ field values, where we expect the runaway in $\phi_Q$ to stop because we know that $m_Q^2 > 0$ in the UV description. To establish consistency with the RGE analysis in the electric theory, where we unambiguously obtain $m_Q^2 > 0$, we make the assumption that $m_Q^2 >0$ in the twice dual theory as well. This leads us to the $\chi$SB minimum in~\Cref{sec:VacuumelectricTwicedual}. The fact that $\chi$SB is also present in the Free Magnetic phase, as noted above, would imply by continuity that the $\chi$SB persists for all $N_f$ in between. 

\begin{figure}[ht]
    \centering
    \includegraphics[width=0.7\linewidth]{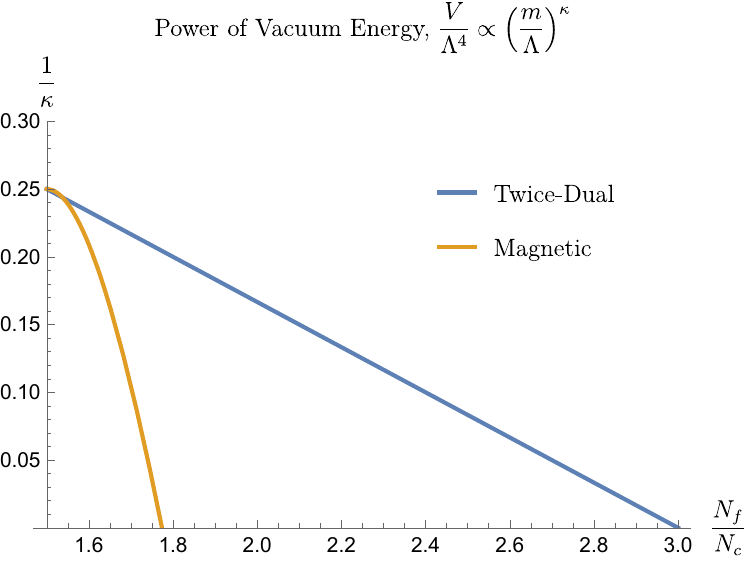}
    \caption{The power of $m/\Lambda$ in the vacuum energy across the conformal window. The result from analyzing the Twice-Dual theory for $N_f/N_c$ near $3$ is in Blue and the result from the Magnetic theory for $N_f/N_c$ near $3/2$ is in Orange. Both results are only valid in the near-edge limit, and are extrapolated to intermediate $N_f/N_c$. It can be seen that the magnetic solution is poorly behaved away from the lower edge, while the Twice-Dual result is surprisingly well-behaved throughout and asymptotes to the same power at $N_f/N_c=3/2$. (Note that the vacuum energy itself, as obtained in the magnetic description, vanishes as $N_f\to 3N_c/2$.)  However, it is important to note that the Twice-Dual result does not reproduce the leading-order quadratic behavior near the bottom edge.}
    \label{fig:vaccumCW}
\end{figure}

For the intermediate range $\frac{3N_c}{2} \ll  N_f \ll 3N_c$, the low-energy theory is strongly coupled and we lack the calculational tools to analyze the vacuum. However, if the $\chi$SB minimum persists in this range, it can be expected to smoothly interpolate between the lower- and upper-edge results. The extrapolation of our near-edge results, in terms of the vacuum energy power law, is shown in~\cref{fig:vaccumCW}. Note that the magnetic dynamical scale $\tilde{\Lambda}$ is related to the electric scale $\Lambda$ by matching at some scale $\mu_m$ via~\cref{eq:matchingscale}. In the plot we take $\mu_m \sim \Lambda$. It should also be noted that as $N_f\to \frac{3N_c}{2}$ the minimum of the magnetic description is driven to zero by its coefficient. SUSY is restored in that limit, in agreement with what one would find starting from $N_f=\frac{3N_c}{2}$ to begin with.

Since the power of the vacuum energy in $m/\Lambda$ diverges as we approach the upper edge of the window, we can deduce that the vacuum energy vanishes as we approach $N_f\to 3N_c$. This may indicate that SUSY is restored in the IR of the AMSB theory for $N_f=3N_c$, since vacuum energy is the order parameter for SUSY breaking. Above $N_f>3N_c$ the AMSB theory has genuine runaways and is not suitable for analyzing the non-SUSY limit. It seems plausible then that there is a turnover point.

Finally, we reiterate that our findings on chiral symmetry breaking appear to be in agreement with lattice results~\cite{DeGrand:2015zxa}. Estimated values, from various approaches, of $N_f$ for the lower edge of the SU(3) conformal window lie in the range 8-13~\cite{Appelquist:1988yc,Cohen:1988sq,Sannino:2004qp,Dietrich:2006cm,Armoni:2009jn,Braun:2009ns,Frandsen:2010ej,Rychkov_2017,Kim:2020yvr,Lee:2020ihn}. Whether $N_f=12$ is in the conformal window or not has attracted recent lattice studies~\cite{Appelquist:2007hu,Appelquist:2009ty,Fodor:2009wk,Fodor:2011tu,Hasenfratz:2011xn,DeGrand:2011cu,Lin:2012iw,Aoki:2012eq,LatKMI:2013bhp,Fodor:2016zil,Hasenfratz:2016dou,Mickley:2025mjj}. Some studies indicate that the chiral symmetry broken ($\chi$SB) vacuum is realized up to and possibly beyond $N_f=3N_c$. For instance, one paper finds that $SU(2)$ gauge theories exhibit chiral symmetry breaking for $N_f < 6$~\cite{Amato:2018nvj}, while another finds that $SU(3)$ gauge theory with $N_f=8$ breaks chiral symmetry~\cite{LatKMI:2014xoh}, and another finds that $SU(3)$ with $N_f=12$ appears to reach an IR fixed point~\cite{LatKMI:2013bhp}. Agreement with direct lattice calculations further supports our conclusions, and lends credence to the picture that the $m\ll \Lambda$ and $m\gg \Lambda$ limits are continuously connected for these cases.

\section{Conclusion}\label{sec:Conclusion}
The main prediction of this paper is that non-supersymmetric QCD leads to chiral symmetry breaking all the way up to $N_f<3N_c$. This prediction assumes that the phase of the dynamics is continuous from the lower edge to the higher edge of the conformal window, where the AMSB effects are shown to be relevant, while we cannot conclusively determine whether the AMSB effects are relevant near the middle of the conformal window where the fixed point dynamics is non-perturbative.

At both the upper edge and lower edge of the conformal window, we needed to make an assumption about the initial conditions of the couplings and RGE flow in order to obtain a QCD-like chiral symmetry breaking minimum. Under this assumption, the twice-dual theory flows to a positive mass squared for the twice-dual squarks and the magnetic theory flows to a positive mass squared for the dual squarks (near the corresponding IR fixed points). As a result, the dynamics is continuous throughout the conformal window. Note that in both theories, the picture with complementary set of initial conditions lead to runaways toward $\mathcal{O}(\Lambda)$ field values, but is also continuous through the window. In fact, it has been shown that for $1.43N_c\lesssim N_f < \frac{3N_c}{2}$ there are similar baryonic runaways towards an uncalculable minimum. 


Our analysis cannot be applied to the range $N_f \geq 3 N_c$, because there the squark mass-squareds become negative and the AMSB theory has a genuine runaway. We also cannot make conclusions about $N_f=\frac{3N_c}{2}$ because of the obstruction of classical conformal invariance in the dynamical superpotential, although continuity would imply chiral-symmetry-breaking there as well.

The remaining issue is whether there is a phase transition as the size of the AMSB is increased from $m \ll \Lambda$ below the dynamical scale $\Lambda$ to $m \gg \Lambda$. The agreement of our result and some lattice result supports this possibility.  It was suggested that an extension of the holomorphy argument may justify the absence of phase transition \cite{Csaki:2021xhi,Csaki:2021aqv}. But further investigation is warranted to understand this question quantitatively.

\section*{Acknowledgments}

We thank Nathaniel Craig, Teruhiko Kawano, Csaba Cs\'aki and Ofri Telem for useful discussions, careful reading of the manuscript and valuable comments. HM also thanks Tetsuya Onogi and Yoshio Kikukawa for useful information about the lattice literature. The work of D.K. was supported by JSPS Kakenhi Grant No.24KJ0613. The work of HM was supported by the Director, Office of Science, Office of High Energy Physics of the U.S. Department of Energy under the Contract No. DE-AC02-05CH11231, by the NSF grant PHY-1915314, by the JSPS Grant-in-Aid for Scientific Research JP20K03942, MEXT Grant-in-Aid for Transformative Research Areas (A) JP20H05850, JP20A203, by World Premier International Research Center Initiative, MEXT, Japan, and Hamamatsu Photonics, K.K.

\appendix

\newpage
\section{More generic choice of the energy scale in the Twice-Dual}\label{sec:genericmu}
In this section, we re-derive the result in the Twice-Dual theory without assuming that the $M,N$ meson VEVs are equal. We note that the following calculation assumes $m_Q^2 > 0$ in the viscinity of the fixed point. This allows us to decouple the quarks $Q,\tilde{Q}$ (in \Cref{sec:TDWODecoupleQ} we show that the same result can be obtained without decoupling), after which we have
 \begin{align}
W&= N_c\Lambda^3 \left( \frac{\det(Y N)}{\Lambda^{N_f}} \right)^{\frac{1}{N_c}}-Y\Lambda N^{ij}M_{ij},\nonumber\\ 
&= N_c\Lambda^3Y^{\frac{N_f}{N_c}} \left( \frac{ \phi_N}{\Lambda} \right)^{\frac{N_f}{N_c}}-N_f Y\Lambda \phi_N\phi_M,  
\end{align}
where meson $M,N$ are taken to be diagonal and full rank, $M=\phi_M\delta_{ij},N=\phi_N\delta_{ij}$. The scalar potential is
\begin{align}
    V =&  \frac{1}{N_fc_M}\abs{\dpd{W}{\phi_M}}^2 + \frac{1}{N_fc_NZ_N}\abs{\dpd{W}{\phi_N}}^2 -\frac{\dot{\gamma}_N}{4} m^2 N_f c_NZ_N \abs{\phi_N}^2 \nonumber\\
    &+m\left(1-\frac{\gamma_N}{2}\right)\left(\phi_N\dpd{W}{\phi_N}+{\rm h.c.}\right) +m\left(\phi_M\dpd{W}{\phi_M}+{\rm h.c.}\right) - 3mW - {\rm h.c.}
\end{align}
Assuming $\phi_N,\phi_M\in\mathbb{R}$, we have
\begin{align}
\frac{V}{\Lambda^4} =& \frac{N_fY^2}{c_M}\left(\frac{\phi_N}{\Lambda}\right)^2-\frac{1}{4}c_N \dot{\gamma}_NN_f\left(\frac{m}{\Lambda}\right)^2Z_N\left(\frac{\phi_N}{\Lambda}\right)^2+\frac{N_f}{c_NZ_N}\left(Y\frac{\phi_M}{\Lambda}-Y^{\frac{N_f}{N_c}} \left(\frac{\phi_N}{\Lambda}\right)^{\frac{N_f}{N_c }-1}\right)^2\nonumber\\
&-\left(6N_c-2\left( 1-\frac{\gamma_N}{2}\right)N_f^2 \right)Y^{\frac{N_f}{N_c}}\frac{m}{\Lambda} \left( \frac{\phi_N}{\Lambda} \right)^{\frac{N_f}{N_c}} 
+2N_f\left(1+\frac{\gamma_N}{2}\right)Y\frac{m}{\Lambda}\frac{\phi_M}{\Lambda}\frac{\phi_N}{\Lambda} 
\end{align}
Now take canonically normalized fields $\phi_N \rightarrow \phi_N/\sqrt{N_f c_N Z_N}$ and $\phi_M \rightarrow \phi_M/\sqrt{N_f c_M}$. Then set $Z_N = \left(\frac{\mu}{\Lambda}\right)^{\gamma_N}$ to get
\begin{align}
    \frac{V}{\Lambda^4} =
&-\frac{\dot{\gamma}_N}{4} \left(\frac{m}{\Lambda}\right)^2 \left(\frac{\phi_N}{\Lambda}\right)^2  \nonumber\\
+&\frac{1}{N_f}\left(\frac{N_f^{\frac{3N_c-N_f}{2N_c}}Y^{\frac{N_f}{N_c}} }{c_N^{\frac{N_f}{N_c}} } \left( \frac{\mu}{\Lambda} \right)^{-\gamma_N\left(\frac{N_f}{N_c}-\frac{1}{2}\right) } \left( \frac{\phi_N}{\Lambda} \right)^{\frac{N_f}{N_c}-1}-\sqrt{\frac{N_f}{c_Nc_M}} Y \left(\frac{\mu}{\Lambda} \right)^{-\frac{\gamma_N}{2} }\frac{\phi_M}{\Lambda}\right)^2\nonumber\\
& +\frac{1}{c_Mc_N}Y^2 \left(\frac{\mu}{\Lambda} \right)^{-\gamma_N}\left( \frac{\phi_N}{\Lambda}\right)^2 
-\frac{\left(6N_c-2\left( 1-\frac{\gamma_N}{2}\right)N_f \right)}{(c_NN_f)^{\frac{N_f}{2N_c}}}  Y^{\frac{N_f}{N_c}}\frac{m}{\Lambda}\left(\frac{\mu}{\Lambda} \right)^{-\frac{\gamma_N}{2}\frac{N_f}{N_c} } \left( \frac{\phi_N}{\Lambda} \right)^{\frac{N_f}{N_c}} \nonumber\\
&+\frac{2}{\sqrt{c_Mc_N}}\left(1+\frac{\gamma_N}{2}\right)Y\frac{m}{\Lambda}\left( \frac{\mu}{\Lambda}\right)^{-\frac{\gamma_N}{2} } \frac{\phi_M}{\Lambda}\frac{\phi_N}{\Lambda}
\end{align}
Take $\gamma_N = -2\frac{3N_c-N_f}{N_f}+c_\gamma\left(\frac{\mu}{\Lambda}\right)^\beta$ and then set $\mu=\phi_M^{p} \phi_N^q$ with $(p+q=1)$ as a fairly generic choice of the energy scale~\cite{Coleman:1973jx} \footnote{The mass dimension of $\mu$ is one. One might wonder if the linear combination of that from $\sum_{k=1}^{n}\phi^{p_k}\phi_N^{q_k},\ (p_k+q_k=1)$ for each $k$ is more generic choice. From the inequality $\sum_{k=1}^{n}\phi^{p_k}\phi_N^{q_k}\geq n\phi^{\frac{1}{n}\sum_{k=1}^{n}p_k}\phi_N^{\frac{1}{n}\sum_{k=1}^{n}q_k}$, the lower bound again is in the form $\phi^{p}\phi_N^{q}\ (p+q=1)$.}. Then
\begin{align}
\frac{V}{\Lambda^4}=
&-\frac{c_\gamma\beta}{4} \left(\frac{m}{\Lambda}\right)^2\left(\frac{\phi_M}{\Lambda}\right)^{p\beta} \left(\frac{\phi_N}{\Lambda}\right)^{2+q\beta}  \nonumber\\
+&\left(\frac{N_f^{\frac{N_c-N_f}{2N_c}}Y^{\frac{N_f}{N_c}} }{c_N^{\frac{N_f}{N_c}} } \left( \frac{\phi_M}{\Lambda}\right)^{2p\frac{3N_c-N_f}{N_f}\left(\frac{N_f}{N_c}-\frac{1}{2}\right) } \left( \frac{\phi_N}{\Lambda} \right)^{\frac{N_f}{N_c}-1+2q\frac{3N_c-N_f}{N_f}\left(\frac{N_f}{N_c}-\frac{1}{2}\right)} \right. \\
&\quad \left. -\sqrt{\frac{1}{c_Nc_M}} Y\left(\frac{\phi_M}{\Lambda}\right)^{1+p\frac{3N_c-N_f}{N_f}}\left(\frac{\phi_N}{\Lambda}\right)^{q\frac{3N_c-N_f}{N_f}}\right)^2\nonumber\\
&+\frac{Y^2}{c_Mc_N}\left( \frac{\phi_M}{\Lambda} \right)^{2p\frac{3N_c-N_f}{N_f}}\left( \frac{\phi_N}{\Lambda}\right)^{2+2q\frac{3N_c-N_f}{N_f}} \nonumber\\
-&\frac{N_fc_\gamma}{(c_NN_f)^{\frac{N_f}{2N_c}}}  Y^{\frac{N_f}{N_c}}\frac{m}{\Lambda}\left(\frac{\phi_M}{\Lambda}\right)^{p\frac{3N_c-N_f}{N_c}+p\beta} \left(\frac{\phi_N}{\Lambda} \right)^{\frac{N_f}{N_c}+q\frac{3N_c-N_f}{N_c}+q\beta} \nonumber\\
&+\frac{2}{\sqrt{c_Mc_N}}\frac{2N_f-3N_c}{N_f}Y\frac{m}{\Lambda} \left(\frac{\phi_M}{\Lambda}\right)^{1+p\frac{3N_c-N_f}{N_f}}\left(\frac{\phi_N}{\Lambda}\right)^{1+q\frac{3N_c-N_f}{N_f}}
    \label{eq:fullpotTWDq}
\end{align}
For the purpose of power counting, consider $N_f=\frac{3N_c}{1+\epsilon}\simeq 3N_c(1-\epsilon)$. Truncating at order $2+\mathcal{O}(\epsilon)$ in the fields gives
\begin{align}\label{eq:potentialdecoupleQ}
\frac{V}{\Lambda^4} \simeq&-\frac{c_\gamma\beta}{4} \left(\frac{m}{\Lambda}\right)^2\left(\frac{\phi_M}{\Lambda}\right)^{p\beta} \left(\frac{\phi_N}{\Lambda}\right)^{2+q\beta}  \nonumber\\
&+\frac{Y^2}{c_Nc_M}\left(\frac{\phi_M}{\Lambda}\right)^{2+2p\frac{3N_c-N_f}{N_f}}\left(\frac{\phi_N}{\Lambda}\right)^{2q\frac{3N_c-N_f}{N_f}}\nonumber\\
&+\frac{Y^2}{c_Mc_N}\left( \frac{\phi_M}{\Lambda} \right)^{2p\frac{3N_c-N_f}{N_f}}\left( \frac{\phi_N}{\Lambda}\right)^{2+2q\frac{3N_c-N_f}{N_f}} \nonumber\\
&+\frac{2}{\sqrt{c_Mc_N}}\frac{2N_f-3N_c}{N_f}Y\frac{m}{\Lambda} \left(\frac{\phi_M}{\Lambda}\right)^{1+p\frac{3N_c-N_f}{N_f}}\left(\frac{\phi_N}{\Lambda}\right)^{1+q\frac{3N_c-N_f}{N_f}}
\end{align}
With $N_f = \frac{3N_c}{1+\epsilon}$, $\beta=21\epsilon^2$ in~\cref{sec:RGEelectricTwiceDual}, and drop terms with coefficients $\mathcal{O}(\epsilon)$,
\begin{align}
V \simeq&
\frac{Y^2}{c_Nc_M}\left(\frac{\phi_M}{\Lambda}\right)^{2+2p\epsilon}\left(\frac{\phi_N}{\Lambda}\right)^{2q\epsilon}
+\frac{Y^2}{c_Mc_N}\left( \frac{\phi_M}{\Lambda} \right)^{2p\epsilon}\left( \frac{\phi_N}{\Lambda}\right)^{2+2q\epsilon} \nonumber\\
&+\frac{2}{\sqrt{c_Mc_N}}Y\frac{m}{\Lambda} \left(\frac{\phi_M}{\Lambda}\right)^{1+p\epsilon}\left(\frac{\phi_N}{\Lambda}\right)^{1+q\epsilon}
\end{align}
With rotation $U(1)_R$, we can rotate a phase of $m$ to take $m\to -m$ in the Weyl compensator (we alternatively could have taken a nonzero phase for $\varphi_M,\varphi_N$). The potential is
\begin{align}
    \frac{V}{\Lambda^4}\simeq& \frac{Y^2}{c_Nc_M}\left(\frac{\phi_M}{\Lambda}\right)^{2+2p\epsilon}\left(\frac{\phi_N}{\Lambda}\right)^{2q\epsilon}
+\frac{Y^2}{c_Mc_N}\left( \frac{\phi_M}{\Lambda} \right)^{2p\epsilon}\left( \frac{\phi_N}{\Lambda}\right)^{2+2q\epsilon} \nonumber\\
&-\frac{2}{\sqrt{c_Mc_N}}Y\frac{m}{\Lambda} \left(\frac{\phi_M}{\Lambda}\right)^{1+p\epsilon}\left(\frac{\phi_N}{\Lambda}\right)^{1+q\epsilon}
\end{align}
This is the potential used in~\cref{eq:TWDmainpotential} with the generic choice of the energy scale $\mu=\phi_M^p\phi_N^q\ (p+q=1)$.\\
The equation of motion is
\begin{align}
    \Lambda\frac{\partial}{\partial\phi_M}\frac{V}{\Lambda^4}
    &=\frac{2Y^2}{c_Nc_M}\left(\frac{\phi_M}{\Lambda}\right)^{1+2p\epsilon}\left(\frac{\phi_N}{\Lambda}\right)^{2q\epsilon}
-\frac{2Y}{\sqrt{c_Mc_N}}\frac{m}{\Lambda} \left(\frac{\phi_M}{\Lambda}\right)^{p\epsilon}\left(\frac{\phi_N}{\Lambda}\right)^{1+q\epsilon}=0,\\
 \Lambda\frac{\partial}{\partial\phi_N}\frac{V}{\Lambda^4}
&=\frac{2Y^2}{c_Mc_N}\left( \frac{\phi_M}{\Lambda} \right)^{2p\epsilon}\left( \frac{\phi_N}{\Lambda}\right)^{1+2q\epsilon} 
-\frac{2Y}{\sqrt{c_Mc_N}}\frac{m}{\Lambda} \left(\frac{\phi_M}{\Lambda}\right)^{1+p\epsilon}\left(\frac{\phi_N}{\Lambda}\right)^{q\epsilon}=0.
\end{align}
The solution of this equation is:
\begin{align}
    \frac{\phi_M}{\Lambda} =\frac{\phi_N}{\Lambda} =& \left(\frac{\sqrt{c_Mc_N}}{Y}\frac{m}{\Lambda}\right)^{\frac{1}{\epsilon}} 
\end{align}

\section{Vacuum of the Twice-Dual theory without Decoupling Quarks}\label{sec:TDWODecoupleQ}
Here we re-derive the same result for the minimum of the Twice-Dual theory without decoupling the quarks.

Start from the same superpotential
\begin{align}
W&=YN(Q\tilde{Q}-M) \rightarrow YN(Q\tilde{Q}-\Lambda M) \nonumber\\
&= YN_fN_c\phi_N\phi_Q\phi_{\tilde{Q}}-YN_{f}\Lambda \phi_N\phi_M
\end{align}
here $Q=\phi_Q \delta^a_{i}, \tilde{Q}=\phi_{\tilde{Q}}\delta^a_i$, $N=\phi_N\delta_{ij}, M=\phi_M\delta_{ij}$ is supposed.

Following the procedure to derive~\cref{eq:AMSBLagrangiantouse} in~\Cref{sec:AMSB}, the scalar potential is
\begin{align}
    V=
&-c_NN_fZ_N(\mu)m^2\frac{\dot{\gamma}_N}{4} |\phi_N|^2
+\frac{1}{c_NN_fZ_N(\mu)} \left| \frac{\partial W}{\partial\phi_N} \right|^2 +\frac{1}{c_MN_f} \left| \frac{\partial W}{\partial\phi_M}\right|^2 \nonumber\\
-&c_QN_fZ_Q(\mu)m^2\frac{\dot{\gamma}_Q}{4} |\phi_Q|^2
-c_{\tilde{Q}}N_fZ_{\tilde{Q}}(\mu)m^2\frac{\dot{\gamma}_{\tilde{Q}}}{4} |\phi_{\tilde{Q}}|^2\nonumber\\
+&\frac{1}{c_QN_fZ_Q(\mu)} \left| \frac{\partial W}{\partial\phi_Q} \right|^2 
+\frac{1}{c_{\tilde{Q}}N_fZ_{\tilde{Q}}(\mu)} \left| \frac{\partial W}{\partial\phi_{\tilde{Q}}} \right|^2  \nonumber\\
-&3m(W+h.c.)
+m \left(\phi_M\frac{\partial W }{\partial \phi_M }+h.c.\right) 
+m \left( 1-\frac{\gamma_N}{2} \right)\left(\phi_N\frac{\partial W }{\partial \phi_N }+h.c.\right) \nonumber\\
+&m \left( 1-\frac{\gamma_Q}{2} \right)\left(\phi_Q\frac{\partial W }{\partial \phi_Q }+h.c.\right) 
+m \left( 1-\frac{\gamma_{\tilde{Q}}}{2} \right)\left(\phi_{\tilde{Q}}\frac{\partial W }{\partial \phi_{\tilde{Q}} }+h.c.\right)
\end{align}
Here,
\begin{align}
\frac{\partial W}{\partial \phi_N} 
&= YN_fN_c\phi_Q\phi_{\tilde{Q}}-N_{f}Y\Lambda \phi_M\\
\frac{\partial W}{\partial \phi_M}
&= -N_{f} Y\Lambda \phi_N \\
\frac{\partial W}{\partial Q}
&=YN_fN_c\phi_N\phi_{\tilde{Q}} \\
\frac{\partial W}{\partial \tilde{Q}} 
&=YN_fN_c\phi_N\phi_Q
\end{align}
Using $Z_N=\left( \frac{\mu}{\Lambda} \right)^{\gamma_N},{Z}_Q=\left( \frac{\mu}{\Lambda} \right)^{\gamma_Q},{Z}_{\tilde{Q}}=\left( \frac{\mu}{\Lambda} \right)^{\gamma_{\bar{Q}}}$, and $\gamma_x = \gamma_x^*+ a_x \left(\frac{\mu}{\Lambda}\right)^{\alpha_x}$, where $\gamma_x^*$ is the value at fixed point, the potential is
\begin{align}
V
=&-c_N a_N N_fm^2\frac{\alpha_N}{4}  \left( \frac{\mu}{\Lambda}\right)^{\alpha_N+\gamma_N} |\phi_N|^2
+\frac{\Lambda^4Y^2}{c_NN_f} \left(\frac{\mu}{\Lambda}\right)^{-\gamma_N} \left| N_fN_c\frac{\phi_Q}{\Lambda}\frac{\phi_{\tilde{Q}}}{\Lambda}-N_f\frac{\phi_M}{\Lambda}\right|^2 \nonumber\\
+&\frac{N_f\Lambda^4}{c_M} Y^2\left| \frac{\phi_N}{\Lambda}\right|^2 
-c_Qa_QN_fm^2\frac{\alpha_Q}{4}  \left( \frac{\mu}{\Lambda}\right)^{\alpha_Q+\gamma_Q} |\phi_Q|^2
-c_{\tilde{Q}}a_{\tilde{Q}}N_fm^2\frac{\alpha_{\tilde{Q}}}{4}  \left( \frac{\mu}{\Lambda}\right)^{\alpha_{\tilde{Q}}+\gamma_{\tilde{Q}}} |\phi_{\tilde{Q}}|^2 \nonumber\\
+&\frac{N_fN_c^2\Lambda^4Y^2}{c_Q} \left(\frac{\phi_N}{\Lambda}\right)^2\left(\frac{\phi_{\tilde{Q}}}{\Lambda}\right)^2+\frac{N_fN_c^2\Lambda^4Y^2}{c_{\tilde{Q}}} \left(\frac{\phi_N}{\Lambda}\right)^2\left(\frac{\phi_{Q}}{\Lambda}\right)^2\nonumber\\
-&6YN_fN_cm\Lambda^3\frac{\phi_N}{\Lambda}\frac{\phi_Q}{\Lambda}\frac{\phi_{\tilde{Q}} }{\Lambda}+6YN_fm\Lambda^3 \frac{\phi_N}{\Lambda}\frac{\phi_M}{\Lambda}\nonumber\\
+&2m\Lambda^3\frac{\phi_M}{\Lambda} \left[ -N_f Y\frac{ \phi_N}{\Lambda} \right] 
+2m\Lambda^3\left( 1-\frac{\gamma_N}{2} \right) \frac{\phi_N}{\Lambda} \left[ YN_fN_c\frac{\phi_Q}{\Lambda}\frac{\phi_{\tilde{Q}}}{\Lambda}-N_fY \frac{\phi_M}{\Lambda}\right]  \nonumber\\
+&2m\Lambda^3\left( 1-\frac{\gamma_Q}{2} \right) \frac{\phi_Q}{\Lambda} \left[ YN_fN_c\frac{\phi_N}{\Lambda}\frac{\phi_{\tilde{Q}}}{\Lambda}\right]
+2m\Lambda^3\left( 1-\frac{\gamma_{\tilde{Q}}}{2} \right) \frac{\phi_{\tilde{Q}}}{\Lambda} \left[ YN_fN_c\frac{\phi_N}{\Lambda}\frac{\phi_Q}{\Lambda}\right]
\end{align}
Here we consider D-flat direction in which $\phi_Q=\phi_{\tilde{Q}}$ (also noting that $\gamma_Q=\gamma_{\tilde{Q}}$, so $a_Q=a_{\tilde{Q}}$ and $\alpha_Q=\alpha_{\tilde{Q}}$)
\begin{align}
\frac{V}{\Lambda^4}
=&-c_Na_N N_f\left(\frac{m}{\Lambda}\right)^2\frac{\alpha_N}{4}  \left( \frac{\mu}{\Lambda}\right)^{\alpha_N+\gamma_N} \left(\frac{\phi_N}{\Lambda}\right)^2
+\frac{N_fY^2}{c_N} \left(\frac{\mu}{\Lambda}\right)^{-\gamma_N} \left| N_c\left(\frac{\phi_Q}{\Lambda}\right)^2-\frac{\phi_M}{\Lambda}\right|^2\nonumber\\ 
+&\frac{N_f}{c_M} Y^2\left| \frac{\phi_N}{\Lambda}\right|^2 
-2c_Qa_QN_f\left(\frac{m}{\Lambda}\right)^2\frac{\alpha_Q}{4}  \left( \frac{\mu}{\Lambda}\right)^{\alpha_Q+\gamma_Q} \left(\frac{\phi_Q}{\Lambda}\right)^2 
+\frac{2N_fN_c^2Y^2}{c_Q} \left(\frac{\phi_N}{\Lambda}\right)^2\left(\frac{\phi_{Q}}{\Lambda}\right)^2\nonumber\\
-&6YN_fN_c\frac{m}{\Lambda}\frac{\phi_N}{\Lambda} \left(\frac{\phi_Q}{\Lambda} \right)^2+6YN_f\frac{m}{\Lambda} \frac{\phi_N}{\Lambda}\frac{\phi_M}{\Lambda}\nonumber\\
-&2YN_f\frac{m}{\Lambda}\frac{\phi_M}{\Lambda} \frac{ \phi_N}{\Lambda} 
+2N_f\frac{m}{\Lambda}\left( 1-\frac{\gamma_N}{2} \right) \frac{\phi_N}{\Lambda} \left[ YN_c\left(\frac{\phi_Q}{\Lambda}\right)^2-Y \frac{\phi_M}{\Lambda}\right]  \nonumber\\
+&4YN_f N_c\frac{m}{\Lambda}\left( 1-\frac{\gamma_Q}{2} \right)\frac{\phi_N}{\Lambda} \left(\frac{\phi_Q}{\Lambda}\right)^2
\end{align}
We can regard this as a quadratic function of $\left( \frac{\phi_Q}{\Lambda} \right)^2$. Note that $\gamma_N^* = -2\frac{3N_c-N_f}{N_f}<0$ and $\gamma_Q^* = \frac{3N_c-N_f}{N_f}>0$. The coefficient of $\left(\frac{\phi_Q}{\Lambda}\right)^2$ is in units of $\Lambda=1$ for simplicity
\begin{align}
    &-\frac{2N_f N_c Y^2}{c_N}\mu^{-\gamma_N}\phi_M - \frac{1}{2} c_Q a_Q N_f m^2 \alpha_Q \mu^{\alpha_Q+\gamma_Q} + \frac{2N_fN_c^2 Y^2}{c_Q}\phi_N^2 - 6Y N_f N_c m \phi_N \nonumber\\
    &+ 2N_f m \left(1-\frac{\gamma_N}{2}\right)\phi_N Y N_c + 4 Y N_fN_c m \left(1-\frac{\gamma_Q}{2}\right)\phi_N \\
    =& 2N_fN_c Y^2 \left(
    \frac{\mu^{-\gamma_N}\phi_M}{c_N}+\frac{\phi_N^2}{c_Q}
    \right) - 2YN_fN_c m \phi_N\left(
    a_Q\mu^{\alpha_Q} + \frac{a_N}{2}\mu^{\alpha_N} 
    \right) 
    -\frac{1}{2}c_Qa_Q N_f m^2\alpha_Q \mu^{\alpha_Q+\gamma_Q} \label{eq:Qcoefficient}
\end{align}
We focus on the very low energy scale $\mu\sim\phi_M\sim\phi_N\ll m$. Then \cref{eq:Qcoefficient} is dominated by the lowest powers of $\phi_N,\mu$. We know that near the top of the conformal window all the $\alpha,\gamma$ are small, so the last term is the dominant one and (again, in units of $\Lambda=1$)
\begin{align}
    V \simeq& 
    \frac{N_fN_c^2Y}{c_N}\mu^{-\gamma_N}\phi_Q^4 
    -a_Q\frac{c_QN_f\alpha_Q}{2}\mu^{\alpha_Q+\gamma_Q}\phi_Q^2
    +
    f(\phi_M,\phi_N,\mu) 
\end{align}
where $f$ is just a placeholder for brevity. There are two possibilities -- either $a_Q>0$ and the minimum is at $\phi_Q\neq 0$ or $a_Q<0$ and the minimum is at $\phi_Q=0$. This is equivalent to the question of whether the squarks get a positive or negative mass-squared from AMSB. The RGE here is identical to that of the magnetic theory with the exchange of $\tilde{N}_c\to N_c$ and renaming of couplings (compare \Cref{sec:RGEmagnetic} to \Cref{sec:RGEelectricTwiceDual}) albeit with the presence of $\mu_{MN}$, which runs to zero and has no notable effect. So $\phi_Q=0$ will only be the case for half of the space of initial conditions. If the squarks get a negative mass-squared contribution from AMSB, the squark VEVs inevitably turn on, leading to a distinct chiral symmetry breaking pattern $SU(N_f) \times SU(N_f) \rightarrow SU(N_c) \times SU(N_{f}-N_{c}) \times SU(N_f)$. However, this minimum would be at $\mathcal{O}(\Lambda)$ field values, where we expect the runaway in $\phi_Q$ to stop becuase we know that $m_Q^2 > 0$ in the UV description. Much like in the magnetic theory, we will assume that the initial conditions lie in such a region of parameter space that $m_Q^2 > 0$ and $\phi_Q=0$. Then the potential can be written as
\begin{align}
\frac{V}{\Lambda^4}
=&2YN_f\frac{m}{\Lambda}\frac{\phi_M}{\Lambda} \frac{ \phi_N}{\Lambda} \left(1+\frac{\gamma_N}{2}\right)-c_Na_N N_f\left(\frac{m}{\Lambda}\right)^2\frac{\alpha_N}{4}  \left( \frac{\mu}{\Lambda}\right)^{\alpha_N+\gamma_N} \left(\frac{\phi_N}{\Lambda}\right)^2
\nonumber\\
+& N_f Y^2 \left(
\frac{1}{c_N}\left(\frac{\mu}{\Lambda}\right)^{-\gamma_N} \abs{\frac{\phi_M}{\Lambda}}^2 + \frac{1}{c_M}\abs{\frac{\phi_N}{\Lambda}}^2
\right)
\end{align}
\begin{align}
\frac{V}{\Lambda^4}
=&-c_Na_N N_f\left(\frac{m}{\Lambda}\right)^2\frac{\alpha_N}{4}  \left( \frac{\mu}{\Lambda}\right)^{\alpha_N+\gamma_N} \left(\frac{\phi_N}{\Lambda}\right)^2\nonumber\\
+&\frac{N_fY^2}{c_N} \left(\frac{\mu}{\Lambda}\right)^{-\gamma_N} \left(\frac{\phi_M}{\Lambda}\right)^2 
+\frac{N_f}{c_M} Y^2\left(\frac{\phi_N}{\Lambda}\right)^2
+2N_f\frac{m}{\Lambda}Y\left(1+\frac{\gamma_N}{2} \right) \frac{\phi_N}{\Lambda}\frac{\phi_M}{\Lambda} 
\end{align}

We use the canonically normalized field $\phi_M\rightarrow \frac{\phi_M}{\sqrt{c_MN_f}}$ and $\phi_N\rightarrow \frac{\phi_N}{c_NN_fZ_N}$, and the potential is
\begin{align}
\frac{V}{\Lambda^4}
=&-a_N\left(\frac{m}{\Lambda}\right)^2\frac{\alpha_N}{4}\left( \frac{\mu}{\Lambda}\right)^{\alpha_N} \left(\frac{\phi_N}{\Lambda}\right)^2\nonumber\\
+&\frac{Y^2}{c_Nc_M} \left(\frac{\mu}{\Lambda}\right)^{-\gamma_N} \left(\frac{\phi_M}{\Lambda}\right)^2 
+\frac{Y^2}{c_Nc_M}\left(\frac{\mu}{\Lambda}\right)^{-\gamma_N}\left(\frac{\phi_N}{\Lambda}\right)^2
+\frac{2Y}{\sqrt{c_Nc_M}}\frac{m}{\Lambda}\left(1+\frac{\gamma_N}{2}\right) \frac{\phi_N}{\Lambda}\frac{\phi_M}{\Lambda}. 
\end{align}
This is identical to~\cref{eq:potentialdecoupleQ}. Therefore, we can obtain the same solution of chiral symmetry broken vacuum in this procedure.
\\ \\ \\

\clearpage

\bibliographystyle{JHEP}
\bibliography{main}

\end{document}